\documentstyle[amssymb,12pt,a4]{article}

\newcommand \beq{\begin{eqnarray}}
\newcommand \eeq{\end{eqnarray}}
\makeatletter
\newcount\@hour\newcount\@minute\chardef\@x10\chardef\@xv60
\def\tcitime{
\def\@time{%
  \@minute\time\@hour\@minute\divide\@hour\@xv
  \ifnum\@hour<\@x 0\fi\the\@hour:%
  \multiply\@hour\@xv\advance\@minute-\@hour
  \ifnum\@minute<\@x 0\fi\the\@minute
  }}%
\@ifundefined{hyperref}{}{}
\@ifundefined{qExtProgCall}{\def\qExtProgCall#1#2#3#4#5#6{\relax}}{}
\def\QCTOpt[#1]#2{%
  \def\QCTOptB{#1}
  \def\QCTOptA{#2}
}
\def\QCTNOpt#1{%
  \def\QCTOptA{#1}
  \let\QCTOptB\empty
}
\def\Qct{%
  \@ifnextchar[{%
    \QCTOpt}{\QCTNOpt}
}
\def\QCBOpt[#1]#2{%
  \def\QCBOptB{#1}
  \def\QCBOptA{#2}
}
\def\QCBNOpt#1{%
  \def\QCBOptA{#1}
  \let\QCBOptB\empty
}
\def\Qcb{%
  \@ifnextchar[{%
    \QCBOpt}{\QCBNOpt}
}
\def\PrepCapArgs{%
  \ifx\QCBOptA\empty
    \ifx\QCTOptA\empty
      {}%
    \else
      \ifx\QCTOptB\empty
        {\QCTOptA}%
      \else
        [\QCTOptB]{\QCTOptA}%
      \fi
    \fi
  \else
    \ifx\QCBOptA\empty
      {}%
    \else
      \ifx\QCBOptB\empty
        {\QCBOptA}%
      \else
        [\QCBOptB]{\QCBOptA}%
      \fi
    \fi
  \fi
}
\newcount\GRAPHICSTYPE
\GRAPHICSTYPE=\z@
\def\GRAPHICSPS#1{%
 \ifcase\GRAPHICSTYPE
   \special{ps: #1}%
 \or
   \special{language "PS", include "#1"}%
 \fi
}%
\def\graffile#1#2#3#4{%
    \leavevmode
    \raise -#4 \BOXTHEFRAME{%
        \hbox to #2{\raise #3\hbox to #2{\null #1\hfil}}}%
}%
\def\draftbox#1#2#3#4{%
 \leavevmode\raise -#4 \hbox{%
  \frame{\rlap{\protect\tiny #1}\hbox to #2%
   {\vrule height#3 width\z@ depth\z@\hfil}%
  }%
 }%
}%
\newcount\draft
\draft=\z@

\newif\ifwasdraft
\wasdraftfalse
\def\GRAPHIC#1#2#3#4#5{%
 \ifnum\draft=\@ne\draftbox{#2}{#3}{#4}{#5}%
  \else\graffile{#1}{#3}{#4}{#5}%
  \fi
 }%
\def\addtoLaTeXparams#1{%
    \edef\LaTeXparams{\LaTeXparams #1}}%
\newif\ifBoxFrame \BoxFramefalse
\newif\ifOverFrame \OverFramefalse
\newif\ifUnderFrame \UnderFramefalse
\def\BOXTHEFRAME#1{%
   \hbox{%
      \ifBoxFrame
         \frame{#1}%
      \else
         {#1}%
      \fi
   }%
}
\def\doFRAMEparams#1{\BoxFramefalse\OverFramefalse\UnderFramefalse\readFRAMEparams#1\end}%
\def\readFRAMEparams#1{%
 \ifx#1\end%
  \let\next=\relax
  \else
  \ifx#1i\dispkind=\z@\fi
  \ifx#1d\dispkind=\@ne\fi
  \ifx#1f\dispkind=\tw@\fi
  \ifx#1t\addtoLaTeXparams{t}\fi
  \ifx#1b\addtoLaTeXparams{b}\fi
  \ifx#1p\addtoLaTeXparams{p}\fi
  \ifx#1h\addtoLaTeXparams{h}\fi
  \ifx#1X\BoxFrametrue\fi
  \ifx#1O\OverFrametrue\fi
  \ifx#1U\UnderFrametrue\fi
  \ifx#1w
    \ifnum\draft=1\wasdrafttrue\else\wasdraftfalse\fi
    \draft=\@ne
  \fi
  \let\next=\readFRAMEparams
  \fi
 \next
 }%
\def\IFRAME#1#2#3#4#5#6{%
      \bgroup
      \let\QCTOptA\empty
      \let\QCTOptB\empty
      \let\QCBOptA\empty
      \let\QCBOptB\empty
      #6%
      \parindent=0pt%
      \leftskip=0pt
      \rightskip=0pt
      \setbox0 = \hbox{\QCBOptA}%
      \@tempdima = #1\relax
      \ifOverFrame
          \typeout{This is not implemented yet}%
          \show\HELP
      \else
         \ifdim\wd0>\@tempdima
            \advance\@tempdima by \@tempdima
            \ifdim\wd0 >\@tempdima
               \textwidth=\@tempdima
               \setbox1 =\vbox{%
                  \noindent\hbox to \@tempdima{\hfill\GRAPHIC{#5}{#4}{#1}{#2}{#3}\hfill}\\%
                  \noindent\hbox to \@tempdima{\parbox[b]{\@tempdima}{\QCBOptA}}%
               }%
               \wd1=\@tempdima
            \else
               \textwidth=\wd0
               \setbox1 =\vbox{%
                 \noindent\hbox to \wd0{\hfill\GRAPHIC{#5}{#4}{#1}{#2}{#3}\hfill}\\%
                 \noindent\hbox{\QCBOptA}%
               }%
               \wd1=\wd0
            \fi
         \else
            \ifdim\wd0>0pt
              \hsize=\@tempdima
              \setbox1 =\vbox{%
                \unskip\GRAPHIC{#5}{#4}{#1}{#2}{0pt}%
                \break
                \unskip\hbox to \@tempdima{\hfill \QCBOptA\hfill}%
              }%
              \wd1=\@tempdima
           \else
              \hsize=\@tempdima
              \setbox1 =\vbox{%
                \unskip\GRAPHIC{#5}{#4}{#1}{#2}{0pt}%
              }%
              \wd1=\@tempdima
           \fi
         \fi
         \@tempdimb=\ht1
         \advance\@tempdimb by \dp1
         \advance\@tempdimb by -#2%
         \advance\@tempdimb by #3%
         \leavevmode
         \raise -\@tempdimb \hbox{\box1}%
      \fi
      \egroup%
}%
\def\DFRAME#1#2#3#4#5{%
 \begin{center}
     \let\QCTOptA\empty
     \let\QCTOptB\empty
     \let\QCBOptA\empty
     \let\QCBOptB\empty
     \ifOverFrame 
        #5\QCTOptA\par
     \fi
     \GRAPHIC{#4}{#3}{#1}{#2}{\z@}
     \ifUnderFrame 
        \nobreak\par #5\QCBOptA
     \fi
 \end{center}%
 }%
\def\FFRAME#1#2#3#4#5#6#7{%
 \begin{figure}[#1]%
  \let\QCTOptA\empty
  \let\QCTOptB\empty
  \let\QCBOptA\empty
  \let\QCBOptB\empty
  \ifOverFrame
    #4
    \ifx\QCTOptA\empty
    \else
      \ifx\QCTOptB\empty
        \caption{\QCTOptA}%
      \else
        \caption[\QCTOptB]{\QCTOptA}%
      \fi
    \fi
    \ifUnderFrame\else
      \label{#5}%
    \fi
  \else
    \UnderFrametrue%
  \fi
  \begin{center}\GRAPHIC{#7}{#6}{#2}{#3}{\z@}\end{center}%
  \ifUnderFrame
    #4
    \ifx\QCBOptA\empty
      \caption{}%
    \else
      \ifx\QCBOptB\empty
        \caption{\QCBOptA}%
      \else
        \caption[\QCBOptB]{\QCBOptA}%
      \fi
    \fi
    \label{#5}%
  \fi
  \end{figure}%
 }%
\newcount\dispkind%
\def\makeactives{
  \catcode`\"=\active
  \catcode`\;=\active
  \catcode`\:=\active
  \catcode`\'=\active
  \catcode`\~=\active
}
\bgroup
   \makeactives
   \gdef\activesoff{%
      \def"{\string"}
      \def;{\string;}
      \def:{\string:}
      \def'{\string'}
      \def~{\string~}
    }
\egroup
\def\FRAME#1#2#3#4#5#6#7#8{%
 \bgroup
 \@ifundefined{bbl@deactivate}{}{\activesoff}
 \ifnum\draft=\@ne
   \wasdrafttrue
 \else
   \wasdraftfalse%
 \fi
 \def\LaTeXparams{}%
 \dispkind=\z@
 \def\LaTeXparams{}%
 \doFRAMEparams{#1}%
 \ifnum\dispkind=\z@\IFRAME{#2}{#3}{#4}{#7}{#8}{#5}\else
  \ifnum\dispkind=\@ne\DFRAME{#2}{#3}{#7}{#8}{#5}\else
   \ifnum\dispkind=\tw@
    \edef\@tempa{\noexpand\FFRAME{\LaTeXparams}}%
    \@tempa{#2}{#3}{#5}{#6}{#7}{#8}%
    \fi
   \fi
  \fi
  \ifwasdraft\draft=1\else\draft=0\fi{}%
  \egroup
 }%
\def\TEXUX#1{"texux"}
\long\def\QQQ#1#2{%
     \long\expandafter\def\csname#1\endcsname{#2}}%
\@ifundefined{QTP}{\def\QTP#1{}}{}
\@ifundefined{QEXCLUDE}{\def\QEXCLUDE#1{}}{}
\@ifundefined{Qlb}{}{}
\@ifundefined{Qlt}{}{}
\long\def\QQA#1#2{}%
\def\QTR#1#2{{\csname#1\endcsname #2}}
\def\EXPAND#1[#2]#3{}%
\def\NOEXPAND#1[#2]#3{}%
\def\LaTeXparent#1{}%
\def\ChildStyles#1{}%
\def\ChildDefaults#1{}%
\def\QTagDef#1#2#3{}%
\@ifundefined{StyleEditBeginDoc}{}{}
\def\QQfnmark#1{\footnotemark}

\@ifundefined{INDEX}{\def\INDEX#1#2{}{}}{}%
\@ifundefined{SUBINDEX}{\def\SUBINDEX#1#2#3{}{}{}}{}%
\@ifundefined{initial}%
   {\def\initial#1{\bigbreak{\raggedright\large\bf #1}\kern 2\p@\penalty3000}}%
   {}%
\@ifundefined{entry}{}{}%
\@ifundefined{primary}{}{}%
\@ifundefined{secondary}{}{}%
\@ifundefined{ZZZ}{}{\makeatletter\input gnuindex.sty\makeatother\makeindex\makeatletter}%
\@ifundefined{abstract}{%
 \def\abstract{%
  \if@twocolumn
   \section*{Abstract (Not appropriate in this style!)}%
   \else \small 
   \begin{center}{\bf Abstract\vspace{-.5em}\vspace{\z@}}\end{center}%
   \quotation 
   \fi
  }%
 }{%
 }%
\@ifundefined{endabstract}{\def\endabstract
  {\if@twocolumn\else\endquotation\fi}}{}%
\@ifundefined{maketitle}{\def\maketitle#1{}}{}%
\@ifundefined{affiliation}{\def\affiliation#1{}}{}%
\@ifundefined{proof}{}{}%
\@ifundefined{endproof}{}{}%
\@ifundefined{newfield}{\def\newfield#1#2{}}{}%
\@ifundefined{chapter}{\def\chapter#1{\par(Chapter head:)#1\par }%
 \newcount\c@chapter}{}%
\@ifundefined{part}{\def\part#1{\par(Part head:)#1\par }}{}%
\@ifundefined{section}{\def\section#1{\par(Section head:)#1\par }}{}%
\@ifundefined{subsection}{\def\subsection#1%
 {\par(Subsection head:)#1\par }}{}%
\@ifundefined{subsubsection}{\def\subsubsection#1%
 {\par(Subsubsection head:)#1\par }}{}%
\@ifundefined{paragraph}{\def\paragraph#1%
 {\par(Subsubsubsection head:)#1\par }}{}%
\@ifundefined{subparagraph}{\def\subparagraph#1%
 {\par(Subsubsubsubsection head:)#1\par }}{}%
\@ifundefined{therefore}{}{}%
\@ifundefined{backepsilon}{}{}%
\@ifundefined{yen}{}{}%
\@ifundefined{registered}{%
   \def\registered{\relax\ifmmode{}\r@gistered
                    \else$\m@th\r@gistered$\fi}%
 \def\r@gistered{^{\ooalign
  {\hfil\raise.07ex\hbox{$\scriptstyle\rm\text{R}$}\hfil\crcr
  \mathhexbox20D}}}}{}%
\@ifundefined{Eth}{}{}%
\@ifundefined{eth}{}{}%
\@ifundefined{Thorn}{}{}%
\@ifundefined{thorn}{}{}%
\def\TEXTsymbol#1{\mbox{$#1$}}%
\@ifundefined{degree}{}{}%
\newdimen\theight
\def\Column{%
 \vadjust{\setbox\z@=\hbox{\scriptsize\quad\quad tcol}%
  \theight=\ht\z@\advance\theight by \dp\z@\advance\theight by \lineskip
  \kern -\theight \vbox to \theight{%
   \rightline{\rlap{\box\z@}}%
   \vss
   }%
  }%
 }%
\def\qed{%
 \ifhmode\unskip\nobreak\fi\ifmmode\ifinner\else\hskip5\p@\fi\fi
 \hbox{\hskip5\p@\vrule width4\p@ height6\p@ depth1.5\p@\hskip\p@}%
 }%
\def\miss{\hbox{\vrule height2\p@ width 2\p@ depth\z@}}%
\def\tcol#1{{\baselineskip=6\p@ \vcenter{#1}} \Column}  %
\def\newfmtname{LaTeX2e}
\def\chkcompat{%
   \if@compatibility
   \else
     \usepackage{latexsym}
   \fi
}
\ifx\fmtname\newfmtname
  \DeclareOldFontCommand{\rm}{\normalfont\rmfamily}{\mathrm}
  \DeclareOldFontCommand{\sf}{\normalfont\sffamily}{\mathsf}
  \DeclareOldFontCommand{\tt}{\normalfont\ttfamily}{\mathtt}
  \DeclareOldFontCommand{\bf}{\normalfont\bfseries}{\mathbf}
  \DeclareOldFontCommand{\it}{\normalfont\itshape}{\mathit}
  \DeclareOldFontCommand{\sl}{\normalfont\slshape}{\@nomath\sl}
  \DeclareOldFontCommand{\sc}{\normalfont\scshape}{\@nomath\sc}
  \chkcompat
\fi
\def\alpha{{\Greekmath 010B}}%
\def\beta{{\Greekmath 010C}}%
\def\gamma{{\Greekmath 010D}}%
\def\delta{{\Greekmath 010E}}%
\def\epsilon{{\Greekmath 010F}}%
\def\zeta{{\Greekmath 0110}}%
\def\eta{{\Greekmath 0111}}%
\def\theta{{\Greekmath 0112}}%
\def\iota{{\Greekmath 0113}}%
\def\kappa{{\Greekmath 0114}}%
\def\lambda{{\Greekmath 0115}}%
\def\mu{{\Greekmath 0116}}%
\def\nu{{\Greekmath 0117}}%
\def\xi{{\Greekmath 0118}}%
\def\pi{{\Greekmath 0119}}%
\def\rho{{\Greekmath 011A}}%
\def\sigma{{\Greekmath 011B}}%
\def\tau{{\Greekmath 011C}}%
\def\upsilon{{\Greekmath 011D}}%
\def\phi{{\Greekmath 011E}}%
\def\chi{{\Greekmath 011F}}%
\def\psi{{\Greekmath 0120}}%
\def\omega{{\Greekmath 0121}}%
\def\varepsilon{{\Greekmath 0122}}%
\def\vartheta{{\Greekmath 0123}}%
\def\varpi{{\Greekmath 0124}}%
\def\varrho{{\Greekmath 0125}}%
\def\varsigma{{\Greekmath 0126}}%
\def\varphi{{\Greekmath 0127}}%
\def\nabla{{\Greekmath 0272}}
\def\FindBoldGroup{%
   {\setbox0=\hbox{$\mathbf{x\global\edef\theboldgroup{\the\mathgroup}}$}}%
}
\def\Greekmath#1#2#3#4{%
    \if@compatibility
        \ifnum\mathgroup=\symbold
           \mathchoice{\mbox{\boldmath$\displaystyle\mathchar"#1#2#3#4$}}%
                      {\mbox{\boldmath$\textstyle\mathchar"#1#2#3#4$}}%
                      {\mbox{\boldmath$\scriptstyle\mathchar"#1#2#3#4$}}%
                      {\mbox{\boldmath$\scriptscriptstyle\mathchar"#1#2#3#4$}}%
        \else
           \mathchar"#1#2#3#4%
        \fi
    \else
        \FindBoldGroup
        \ifnum\mathgroup=\theboldgroup 
           \mathchoice{\mbox{\boldmath$\displaystyle\mathchar"#1#2#3#4$}}%
                      {\mbox{\boldmath$\textstyle\mathchar"#1#2#3#4$}}%
                      {\mbox{\boldmath$\scriptstyle\mathchar"#1#2#3#4$}}%
                      {\mbox{\boldmath$\scriptscriptstyle\mathchar"#1#2#3#4$}}%
        \else
           \mathchar"#1#2#3#4%
        \fi     	    
	  \fi}
\newif\ifGreekBold  \GreekBoldfalse
\let\SAVEPBF=\pbf
\def\pbf{\GreekBoldtrue\SAVEPBF}%
\@ifundefined{theorem}{}{}
\@ifundefined{lemma}{}{}
\@ifundefined{corollary}{}{}
\@ifundefined{conjecture}{}{}
\@ifundefined{proposition}{}{}
\@ifundefined{axiom}{}{}
\@ifundefined{remark}{}{}
\@ifundefined{example}{}{}
\@ifundefined{exercise}{}{}
\@ifundefined{definition}{}{}
\@ifundefined{mathletters}{%
  \newcounter{equationnumber}  
  \def\mathletters{%
     \addtocounter{equation}{1}
     \edef\@currentlabel{\theequation}%
     \setcounter{equationnumber}{\c@equation}
     \setcounter{equation}{0}%
     \edef\theequation{\@currentlabel\noexpand\alph{equation}}%
  }
  
}{}
\@ifundefined{BibTeX}{%
    \def\BibTeX{{\rm B\kern-.05em{\sc i\kern-.025em b}\kern-.08em
                 T\kern-.1667em\lower.7ex\hbox{E}\kern-.125emX}}}{}%
\@ifundefined{AmS}%
    {\def\AmS{{\protect\usefont{OMS}{cmsy}{m}{n}%
                A\kern-.1667em\lower.5ex\hbox{M}\kern-.125emS}}}{}%
\@ifundefined{AmSTeX}{}{}%
\ifx\ds@amstex\relax
\makeatother 
\else
   \@ifpackageloaded{amstex}%
      {\message{amstex already loaded}\makeatother }
      {}
   \@ifpackageloaded{amsgen}%
      {\message{amsgen already loaded}\makeatother }
      {}
\fi
\def\DN@{\def\next@}%
\def\eat@#1{}%
\let\DOTSI\relax
\def\RIfM@{\relax\ifmmode}%
\def\FN@{\futurelet\next}%
\newcount\intno@
\def\iint{\DOTSI\intno@\tw@\FN@\ints@}%
\def\iiint{\DOTSI\intno@\thr@@\FN@\ints@}%
\def\iiiint{\DOTSI\intno@4 \FN@\ints@}%
\def\idotsint{\DOTSI\intno@\z@\FN@\ints@}%
\def\ints@{\findlimits@\ints@@}%
\newif\iflimtoken@
\newif\iflimits@
\def\findlimits@{\limtoken@true\ifx\next\limits\limits@true
 \else\ifx\next\nolimits\limits@false\else
 \limtoken@false\ifx\ilimits@\nolimits\limits@false\else
 \ifinner\limits@false\else\limits@true\fi\fi\fi\fi}%
\def\multint@{\int\ifnum\intno@=\z@\intdots@                          
 \else\intkern@\fi                                                    
 \ifnum\intno@>\tw@\int\intkern@\fi                                   
 \ifnum\intno@>\thr@@\int\intkern@\fi                                 
 \int}
\def\multintlimits@{\intop\ifnum\intno@=\z@\intdots@\else\intkern@\fi
 \ifnum\intno@>\tw@\intop\intkern@\fi
 \ifnum\intno@>\thr@@\intop\intkern@\fi\intop}%
\def\intic@{%
    \mathchoice{\hskip.5em}{\hskip.4em}{\hskip.4em}{\hskip.4em}}%
\def\negintic@{\mathchoice
 {\hskip-.5em}{\hskip-.4em}{\hskip-.4em}{\hskip-.4em}}%
\def\ints@@{\iflimtoken@                                              
 \def\ints@@@{\iflimits@\negintic@
   \mathop{\intic@\multintlimits@}\limits                             
  \else\multint@\nolimits\fi                                          
  \eat@}
 \else                                                                
 \def\ints@@@{\iflimits@\negintic@
  \mathop{\intic@\multintlimits@}\limits\else
  \multint@\nolimits\fi}\fi\ints@@@}%
\def\intkern@{\mathchoice{\!\!\!}{\!\!}{\!\!}{\!\!}}%
\def\plaincdots@{\mathinner{\cdotp\cdotp\cdotp}}%
\def\intdots@{\mathchoice{\plaincdots@}%
 {{\cdotp}\mkern1.5mu{\cdotp}\mkern1.5mu{\cdotp}}%
 {{\cdotp}\mkern1mu{\cdotp}\mkern1mu{\cdotp}}%
 {{\cdotp}\mkern1mu{\cdotp}\mkern1mu{\cdotp}}}%
\def\RIfM@{\relax\protect\ifmmode}
\def\text{\RIfM@\expandafter\text@\else\expandafter\mbox\fi}
\let\nfss@text\text
\def\text@#1{\mathchoice
   {\textdef@\displaystyle\f@size{#1}}%
   {\textdef@\textstyle\tf@size{\firstchoice@false #1}}%
   {\textdef@\textstyle\sf@size{\firstchoice@false #1}}%
   {\textdef@\textstyle \ssf@size{\firstchoice@false #1}}%
   \glb@settings}
\def\textdef@#1#2#3{\hbox{{%
                    \everymath{#1}%
                    \let\f@size#2\selectfont
                    #3}}}
\newif\iffirstchoice@
\firstchoice@true
\def\Let@{\relax\iffalse{\fi\let\\=\cr\iffalse}\fi}%
\def\vspace@{\def\vspace##1{\crcr\noalign{\vskip##1\relax}}}%
\def\multilimits@{\bgroup\vspace@\Let@
 \baselineskip\fontdimen10 \scriptfont\tw@
 \advance\baselineskip\fontdimen12 \scriptfont\tw@
 \lineskip\thr@@\fontdimen8 \scriptfont\thr@@
 \lineskiplimit\lineskip
 \vbox\bgroup\ialign\bgroup\hfil$\m@th\scriptstyle{##}$\hfil\crcr}%
\def\Sb{_\multilimits@}%
\def\endSb{\crcr\egroup\egroup\egroup}%
\def\Sp{^\multilimits@}%

\newdimen\ex@
\def\rightarrowfill@#1{$#1\m@th\mathord-\mkern-6mu\cleaders
 \hbox{$#1\mkern-2mu\mathord-\mkern-2mu$}\hfill
 \mkern-6mu\mathord\rightarrow$}%
\def\leftarrowfill@#1{$#1\m@th\mathord\leftarrow\mkern-6mu\cleaders
 \hbox{$#1\mkern-2mu\mathord-\mkern-2mu$}\hfill\mkern-6mu\mathord-$}%
\def\leftrightarrowfill@#1{$#1\m@th\mathord\leftarrow
\mkern-6mu\cleaders
 \hbox{$#1\mkern-2mu\mathord-\mkern-2mu$}\hfill
 \mkern-6mu\mathord\rightarrow$}%
\def\overrightarrow{\mathpalette\overrightarrow@}%
\def\overrightarrow@#1#2{\vbox{\ialign{##\crcr\rightarrowfill@#1\crcr
 \noalign{\kern-\ex@\nointerlineskip}$\m@th\hfil#1#2\hfil$\crcr}}}%

\def\overleftarrow{\mathpalette\overleftarrow@}%
\def\overleftarrow@#1#2{\vbox{\ialign{##\crcr\leftarrowfill@#1\crcr
 \noalign{\kern-\ex@\nointerlineskip}$\m@th\hfil#1#2\hfil$\crcr}}}%
\def\overleftrightarrow{\mathpalette\overleftrightarrow@}%
\def\overleftrightarrow@#1#2{\vbox{\ialign{##\crcr
   \leftrightarrowfill@#1\crcr
 \noalign{\kern-\ex@\nointerlineskip}$\m@th\hfil#1#2\hfil$\crcr}}}%
\def\underrightarrow{\mathpalette\underrightarrow@}%
\def\underrightarrow@#1#2{\vtop{\ialign{##\crcr$\m@th\hfil#1#2\hfil
  $\crcr\noalign{\nointerlineskip}\rightarrowfill@#1\crcr}}}%

\def\underleftarrow{\mathpalette\underleftarrow@}%
\def\underleftarrow@#1#2{\vtop{\ialign{##\crcr$\m@th\hfil#1#2\hfil
  $\crcr\noalign{\nointerlineskip}\leftarrowfill@#1\crcr}}}%
\def\underleftrightarrow{\mathpalette\underleftrightarrow@}%
\def\underleftrightarrow@#1#2{\vtop{\ialign{##\crcr$\m@th
  \hfil#1#2\hfil$\crcr
 \noalign{\nointerlineskip}\leftrightarrowfill@#1\crcr}}}%
\def\qopnamewl@#1{\mathop{\operator@font#1}\nlimits@}
\let\nlimits@\displaylimits
\def\setboxz@h{\setbox\z@\hbox}
\def\varlim@#1#2{\mathop{\vtop{\ialign{##\crcr
 \hfil$#1\m@th\operator@font lim$\hfil\crcr
 \noalign{\nointerlineskip}#2#1\crcr
 \noalign{\nointerlineskip\kern-\ex@}\crcr}}}}
 \def\rightarrowfill@#1{\m@th\setboxz@h{$#1-$}\ht\z@\z@
  $#1\copy\z@\mkern-6mu\cleaders
  \hbox{$#1\mkern-2mu\box\z@\mkern-2mu$}\hfill
  \mkern-6mu\mathord\rightarrow$}
\def\leftarrowfill@#1{\m@th\setboxz@h{$#1-$}\ht\z@\z@
  $#1\mathord\leftarrow\mkern-6mu\cleaders
  \hbox{$#1\mkern-2mu\copy\z@\mkern-2mu$}\hfill
  \mkern-6mu\box\z@$}
\def\projlim{\qopnamewl@{proj\,lim}}
\def\injlim{\qopnamewl@{inj\,lim}}
\def\varinjlim{\mathpalette\varlim@\rightarrowfill@}
\def\varprojlim{\mathpalette\varlim@\leftarrowfill@}
\def\varliminf{\mathpalette\varliminf@{}}
\def\varliminf@#1{\mathop{\underline{\vrule\@depth.2\ex@\@width\z@
   \hbox{$#1\m@th\operator@font lim$}}}}
\def\varlimsup{\mathpalette\varlimsup@{}}
\def\varlimsup@#1{\mathop{\overline
  {\hbox{$#1\m@th\operator@font lim$}}}}
\def\dfrac#1#2{{\displaystyle {#1 \over #2}}}%
\def\QDOVERD#1#2#3#4{{\displaystyle {#3 \overwithdelims#1#2 #4}}}%
\begingroup \catcode `|=0 \catcode `[= 1
\catcode`]=2 \catcode `\{=12 \catcode `\}=12
\catcode`\\=12 
|gdef|@alignverbatim#1\end{align}[#1|end[align]]
|gdef|@salignverbatim#1\end{align*}[#1|end[align*]]
|gdef|@alignatverbatim#1\end{alignat}[#1|end[alignat]]
|gdef|@salignatverbatim#1\end{alignat*}[#1|end[alignat*]]
|gdef|@xalignatverbatim#1\end{xalignat}[#1|end[xalignat]]
|gdef|@sxalignatverbatim#1\end{xalignat*}[#1|end[xalignat*]]
|gdef|@gatherverbatim#1\end{gather}[#1|end[gather]]
|gdef|@sgatherverbatim#1\end{gather*}[#1|end[gather*]]
|gdef|@gatherverbatim#1\end{gather}[#1|end[gather]]
|gdef|@sgatherverbatim#1\end{gather*}[#1|end[gather*]]
|gdef|@multilineverbatim#1\end{multiline}[#1|end[multiline]]
|gdef|@smultilineverbatim#1\end{multiline*}[#1|end[multiline*]]
|gdef|@arraxverbatim#1\end{arrax}[#1|end[arrax]]
|gdef|@sarraxverbatim#1\end{arrax*}[#1|end[arrax*]]
|gdef|@tabulaxverbatim#1\end{tabulax}[#1|end[tabulax]]
|gdef|@stabulaxverbatim#1\end{tabulax*}[#1|end[tabulax*]]
|endgroup
\def\align{\@verbatim \frenchspacing\@vobeyspaces \@alignverbatim
You are using the "align" environment in a style in which it is not defined.}

\@namedef{align*}{\@verbatim\@salignverbatim
You are using the "align*" environment in a style in which it is not defined.}
\expandafter\let\csname endalign*\endcsname =\endtrivlist
\def\alignat{\@verbatim \frenchspacing\@vobeyspaces \@alignatverbatim
You are using the "alignat" environment in a style in which it is not defined.}

\@namedef{alignat*}{\@verbatim\@salignatverbatim
You are using the "alignat*" environment in a style in which it is not defined.}
\expandafter\let\csname endalignat*\endcsname =\endtrivlist
\def\xalignat{\@verbatim \frenchspacing\@vobeyspaces \@xalignatverbatim
You are using the "xalignat" environment in a style in which it is not defined.}

\@namedef{xalignat*}{\@verbatim\@sxalignatverbatim
You are using the "xalignat*" environment in a style in which it is not defined.}
\expandafter\let\csname endxalignat*\endcsname =\endtrivlist
\def\gather{\@verbatim \frenchspacing\@vobeyspaces \@gatherverbatim
You are using the "gather" environment in a style in which it is not defined.}

\@namedef{gather*}{\@verbatim\@sgatherverbatim
You are using the "gather*" environment in a style in which it is not defined.}
\expandafter\let\csname endgather*\endcsname =\endtrivlist
\def\multiline{\@verbatim \frenchspacing\@vobeyspaces \@multilineverbatim
You are using the "multiline" environment in a style in which it is not defined.}

\@namedef{multiline*}{\@verbatim\@smultilineverbatim
You are using the "multiline*" environment in a style in which it is not defined.}
\expandafter\let\csname endmultiline*\endcsname =\endtrivlist
\def\arrax{\@verbatim \frenchspacing\@vobeyspaces \@arraxverbatim
You are using a type of "array" construct that is only allowed in AmS-LaTeX.}

\def\tabulax{\@verbatim \frenchspacing\@vobeyspaces \@tabulaxverbatim
You are using a type of "tabular" construct that is only allowed in AmS-LaTeX.}

\@namedef{arrax*}{\@verbatim\@sarraxverbatim
You are using a type of "array*" construct that is only allowed in AmS-LaTeX.}
\expandafter\let\csname endarrax*\endcsname =\endtrivlist
\@namedef{tabulax*}{\@verbatim\@stabulaxverbatim
You are using a type of "tabular*" construct that is only allowed in AmS-LaTeX.}
\expandafter\let\csname endtabulax*\endcsname =\endtrivlist
\def\@@eqncr{\let\@tempa\relax
    \ifcase\@eqcnt \def\@tempa{& & &}\or \def\@tempa{& &}%
      \else \def\@tempa{&}\fi
     \@tempa
     \if@eqnsw
        \iftag@
           \@taggnum
        \else
           \@eqnnum\stepcounter{equation}%
        \fi
     \fi
     \global\tag@false
     \global\@eqnswtrue
     \global\@eqcnt\z@\cr}
 \def\endequation{%
     \ifmmode\ifinner 
      \iftag@
        \addtocounter{equation}{-1} 
        $\hfil
           \displaywidth\linewidth\@taggnum\egroup \endtrivlist
        \global\tag@false
        \global\@ignoretrue   
      \else
        $\hfil
           \displaywidth\linewidth\@eqnnum\egroup \endtrivlist
        \global\tag@false
        \global\@ignoretrue 
      \fi
     \else   
      \iftag@
        \addtocounter{equation}{-1} 
        \eqno \hbox{\@taggnum}
        \global\tag@false%
        $$\global\@ignoretrue
      \else
        \eqno \hbox{\@eqnnum}
        $$\global\@ignoretrue
      \fi
     \fi\fi
 } 
 \newif\iftag@ \tag@false
 \def\tag{\@ifnextchar*{\@tagstar}{\@tag}}
 \def\@tag#1{%
     \global\tag@true
     \global\def\@taggnum{(#1)}}
 \def\@tagstar*#1{%
     \global\tag@true
     \global\def\@taggnum{#1}%
}
\makeatother

\begin{document}

\title{{\sf Formation of an ordered phase in neutron star matter}}
\author{M.A. P\'{e}rez Garc\'{\i }a $^{1}\thanks{%
E-mail addresses: aperez@pinon.ccu.uniovi.es, diaz@obspm.fr,
corte@pinon.ccu.uniovi.es, lysiane@fisi24.ciencias.uniovi.es,
gravity@pinon.ccu.uniovi.es}$, J. D\'{\i }az Alonso$^{1,2}$ , N. Corte Rodr{%
\'{\i }}guez $^{1}$, \and L. Mornas $^{1}$, J.P. Su\'{a}rez Curieses $^{1}$ 
\\
{\small {\it (1) Dpto. de F{\'\i}sica, Universidad de Oviedo. Avda. Calvo
Sotelo 18, E-33007 Oviedo, Asturias, SPAIN}}\\
{\small {\it (2) Observatoire de Paris, D.A.R.C. (U.M.R. 8629 C.N.R.S.)
F-92190 Meudon, FRANCE}} \\
}
\maketitle

\begin{abstract}
In this work, we explore the possible formation of ordered phases in
hadronic matter, related to the presence of hyperons at high densities. We
analyze a microscopic mechanism which can lead to the crystallization of the
hyperonic sector by the confinement of the hyperons on the nodes of a
lattice. For this purpose, we introduce a simplified model of the hadronic
plasma, in which the nuclear interaction between protons, neutrons and
hyperons is mediated by meson fields. We find that, for some reasonable sets
of values of the model parameters, such ordered phases are energetically
favoured as density increases beyond a threshold value.
\end{abstract}

\noindent {\small PACS: }

\noindent {\small keywords: hyperons, crystallization, neutron stars,
relativistic nuclear matter}

\section{Introduction}

The formation of solid cores in neutron stars or the presence of solid
phases in hadronic matter has been considered by many authors, invoking
several mechanisms. Whereas the idea of solidification of pure neutron
matter considered in the seventies \cite{seventies} was abandoned, other
mechanisms have received attention lately: pion condensation \cite{japoneses}%
, confinement of protons \cite{kutschera} by the repulsive character of the
in-medium n-p interaction, ordering of a mixed quark-hadron phase in a way
reminiscent of the neutron drip phase in the surface layers \cite
{Glendenning}, {\it etc}. (See also \cite{migdal,lovas,shepard}.)

A solid phase may have interesting observable consequences on the behavior
of neutron stars \cite{d83,haensel97}. As a rotating star slows down, or in
the course of the cooling process, a small change in the star internal
pressure may trigger a sudden change of the central density and be
observable as a discontinuity in the braking index, {\it i.e.} a glitch \cite
{brakingindex}. Similarly, a mini-collapse following a phase transition to
an ordered phase was suggested as a mechanism for anomalous gamma-ray bursts 
\cite{muto}. Moreover, if a sizeable fraction of the matter of the neutron
star interior exists in the solid state, the star supports elastic strain
and develops triaxiality, leading to gravitational wave emission and
precession \cite{haensel97}. Finally, consequences are to be expected on the
mechanism of neutrino emission and neutron star cooling \cite
{haenselcool,prakashcool}.

The equation of state of hadronic matter accounting for the presence of
nucleons and hyperons which interact through the exchange of several mesons
has been extensively studied using phenomenological lagrangian models \cite
{lattimer}. Such models suggest that the presence of hyperons in hadronic
matter starts at a threshold density around 1.5 times nuclear saturation and
increases beyond.

In the mean field approximation to the solution of these models, the
hyperonic sector in the ground state is supposed to form (as the other
baryonic components) an uniformly distributed Fermi fluid undergoing the
action of the mesonic mean fields created by the whole baryonic distribution
in the plasma. As the abundance of hyperons strongly increases with density,
the contribution of this sector to the total energy of the ground state
becomes very important in a Fermi fluid configuration. In these conditions,
one may ask whether such a fluid is stable at high densities, or would other
ground state configurations be energetically more favorable.

In this work, we shall investigate the relative stability of a new class of
ordered configurations for the ground state of hadronic matter, where an
hyperon sector is confined on the nodes of a periodic lattice. For this
purpose, we introduce a model of the hadronic plasma where a system of
baryons (protons, neutrons and hyperons) interact through the exchange of
meson fields in the framework of a relativistic phenomenological lagrangian
theory. The mesonic and hyperonic sectors included in the lagrangian
determine the complexity of the corresponding version of the model, as well
as the accuracy of its predictions. A satisfactory model should include the
whole family of hyperons and the set of mesons allowing a realistic
description of the strong interaction in the relevant range of densities and
energies. Nevertheless, in order to elaborate the techniques for the
description of eventual ordered configurations and the analysis of their
stability, avoiding the complexities related to the microscopic dynamics, we
shall treat here a simplified version, where the interaction is described
through the exchange of scalar ($\sigma $) and vector ($\omega $) mesons.
The assumed baryonic components are protons, neutrons, and we consider only
one species of electrically neutral ($\Lambda $) hyperon coupled to the
mesons, as well as an electronic sector compensating the proton electric
charge.

Although this simple version cannot account for all the features of the
nuclear interaction, when the model is solved in the mean field
approximation with the $\beta $ equilibrium conditions, it allows the
fitting of nuclear saturation properties at nuclear density $n_{0},$ and
leads to a behavior of the thermodynamic functions and baryonic abundances
beyond saturation density similar to what would be obtained from a more
elaborate description. Consequently, it will be a useful laboratory in order
to investigate the lpossible existence of the ordered ground state
configurations, their structure and stability.

In this context, we analyze a configuration where the hyperons in the ground
state of the plasma are confined on the nodes of a lattice, and inquire
whether it is energetically favored, for some density ranges, with respect
to the usual Fermi liquid configuration. In this ordered phase, the hyperons
are treated as harmonic oscillators (gaussian clouds). They are assumed to
be confined around the nodes of a cubic lattice by the potential
self-consistently generated by the other hyperons in the lattice and the
fluid of nucleons.

In order to avoid the complexities related to the spin-spin component of the
nuclear interaction, we analyze a particular configuration where two
hyperons with antiparallel spins coexist in the ground state of the harmonic
potential at every lattice site, where they behave as a bound state of
hyperons. In a first approximation we neglect the interaction energy between
the two hyperons at the same site. The presence of bound states of hyperons
(H-dibaryons \cite{jaffe77}) in hadronic matter at high densities has been
considered in reference \cite{GLEN98} \cite{faess}. Their stability could be
enhanced if the existence of an additional attractive $\Lambda \Lambda $
interaction should be confirmed, as the excess binding of double hypernuclei
seems to indicate \cite{caro}. If this is not so, the interaction between
the two independent hyperons at a node can be obtained as the potential
energy between two gaussian hyperonic clouds (of the form self-consistently
obtained from the lattice potential) centered on the same node. We shall see
that, at least in the relevant range of densities where the solid phase is
found, the balance between the contributions coming from the attractive
spin-spin component of the nuclear potential and the short-range repulsive
central component leads to an interaction energy between the two hyperons at
a node which is attractive (and, consequently, improves the confinement) up
to densities $\sim 3n_{0}$. Beyond this density it becomes repulsive, but
small as compared with the depth of the confining lattice potential, and the
associated corrections can be treated in a perturbative way.

The hyperons confined around the nodes of a regular lattice must induce a
redistribution of the surrounding nucleons. Such a redistribution can be
calculated in the random phase approximation (RPA) \cite{DP91}, by treating
the lattice as a system of impurities perturbing the ambient mean field and
the nucleon Fermi distribution or, more exactly, by solving the Dirac
equation in the periodic potential of the lattice through the Bloch method 
\cite{AS76}. Nevertheless, we shall adopt here a simpler approximation to
the dynamics of the nucleons in the medium, by neglecting the spatial
variations of the lattice potential and treating the nucleon as a
relativistic fermion moving in the constant mean field generated by the
uniform Fermi fluid of nucleons plus the spatial average of the hyperonic
distribution. This approximation is the analog of the Sommerfeld's free
electron approximation in a Coulomb lattice \cite{AS76}. The corrections
introduced by the non-uniform character of the lattice field will be
considered elsewhere.

\section{The model and the solid phase}

We describe the baryon interaction through the exchange of scalar $\sigma $
mesons and vector $\omega $ mesons as in Walecka model \cite{wal}. A
leptonic sector reduced to the electrons is included in order to preserve
the electric charge neutrality. The lagrangian density defining the nucleon
and hyperon coupling to the mesons is

\begin{eqnarray}
L &=&\sum_{B=N,h}\frac{i}{2}[\overline{\Psi _{B}}\gamma \partial \Psi
_{B}-\partial \overline{\Psi _{B}}\gamma \Psi _{B}]-m_{B}\overline{\Psi _{B}}%
\Psi _{B}+g_{\sigma B}\overline{\Psi _{B}}\sigma \Psi _{B}+g_{\omega B}%
\overline{\Psi _{B}}\gamma ^{\mu }\omega _{\mu }\Psi _{B}  \nonumber \\
&&+\frac{1}{2}(\partial _{\mu }\sigma \partial ^{\mu }\sigma -m_{\sigma
}^{2}\sigma ^{2})-\frac{1}{2}(\frac{1}{2}F_{\omega }{}_{\mu \nu }F_{\omega
}{}^{\mu \nu }-m_{\omega }^{2}\omega ^{\nu }\omega _{\nu })
\end{eqnarray}
where the stregths of the couplings between the nucleons and hyperons to the
meson fields are different. This lagrangian is the simplest one which allows
an acceptable fit of nuclear matter saturation properties.

We assume now a configuration in which the hyperons are confined on the
nodes of a cubic lattice by the interaction with the nucleon background and
the fields generated by the lattice itself. If the resulting potential in a
lattice site, $\overrightarrow{{r}_{i}},$ is approximated by an harmonic
potential, the state of each hyperon can be described as the ground state of
a classical harmonic oscillator characterized by a gaussian wave function of
the form ($\hslash =c=1)$

\begin{equation}
\Psi _{h}(\overrightarrow{{r}})=\left( {\frac{b}{\pi }}\right) ^{(3/4)}\exp
(-\frac{b}{2}(\overrightarrow{{r}}-\overrightarrow{{r}_{i}})^{2})\qquad
;\qquad b=M_{h}\,\nu _{0}
\end{equation}
\label{eq:(2.2)}where $M_{h}$ is the effective hyperon mass, that is the
free mass modified by the mean scalar field generated by the uniform nucleon
distribution (see (\ref{mh}) below). The oscillator frequency $\nu _{0}$ is
related to the central curvature of the harmonic potential, taken as the
parabolic potential osculating the true lattice potential on the node. This
lattice potential at every site is obtained as the superposition of the
fields created by the gaussian clouds of the hyperons in the other sites.

As mentioned in the introduction, we shall consider the configuration where
two hyperons with antiparallel spins fill the ground state of the harmonic
potential at every site. In this way the spin-spin interaction contribution
to the lattice potential which comes from every node has the same form,
avoiding the difficulties related to a configuration of individual hyperons
with a random or complex distribution of spins. Moreover, the dynamics of
hyperons in these states can be treated as non-relativistic. Relativistic
corrections will be added in a later stage in the calculation of the
thermodynamic state. This is consistent with the treatment of the dynamics
of the other baryons in the ordered phase and with the calculation of the
Fermi liquid configuration, to which one must compare in analyzing the
relative stability of the ground states.

As a first step, we calculate the field created by a gaussian cloud at a
given distance of the center. In the present case is given by the potential

\begin{equation}
V(\overrightarrow{r})=\int d_{3}\,\overrightarrow{q}\ \exp (-i%
\overrightarrow{q}.\overrightarrow{r})\left[ g_{\sigma h}\frac{F_{\sigma
}(q)S_{h}(q)}{(q^{2}+m_{\sigma }^{2})}+g_{\omega h}\frac{F_{\omega
}(q)n_{h}(q)}{(q^{2}+m_{\omega }^{2})}\right]  \label{v}
\end{equation}
\label{eq:(2.3)}where $S_{h}(q)$ and $n_{h}(q)$ are the hyperonic scalar and
particle density associated to the gaussian cloud in the $q$-space\footnote{%
in the non relativistic treatement, the hyperonic density $n_{h}$ and scalar
density $S_{h}$, which are the sources of the vector and scalar fields
respectively, coincide.}. We introduced form factors as usual in order to
take into account the fact that the hyperons are composite extended particles

\begin{equation}
F_{i}(q)=\frac{\Lambda _{i}^{2}-m_{i}^{2}}{\Lambda _{i}^{2}+q^{2}}\qquad
;\qquad i=\sigma ,\omega
\end{equation}
\label{eq:(2.4)}From this expression, we can obtain the potential energy of
a point-like hyperon of given spin, at a point $\vec{r}$ in the total field
created by the gaussian cloud associated to the two antiparallel-spin
hyperons centered at $\vec{r}\ =\ 0$:

\begin{equation}
U(\overrightarrow{r})=\frac{2\pi }{r}\int_{0}^{\infty
}Rn(R)dR\int_{|r-R|}^{|r+R|}x\Phi (x)dx  \label{pot}
\end{equation}
\label{eq:(2.5)}where $n(r)$ is the gaussian charge distribution of the two
hyperons in the node, given by

\begin{equation}
n(r)=2\Psi _{h}^{*}(r)\Psi _{h}(r)=2\left( \frac{M_{h}\nu _{0}}{\pi }\right)
^{3/2}e^{-M_{h}\nu _{0}r^{2}}
\end{equation}
\label{eq:(2.7)}and $\Phi (x)$ is the expression of the elementary potential
energy between two hyperon states in the one-boson exchange approximation.
At zeroth order in momentum expansion and excluding the effects of the form
factors, $\Phi (x)$ reduces to the central component only, which has the
usual Yukawa expression for both meson exchanges:

\begin{equation}
\Phi (x)=-\frac{g_{\sigma h}^{2}}{4\pi }\frac{e^{-m_{\sigma }r}}{r}+\frac{%
g_{\omega h}^{2}}{4\pi }^{2}\frac{e^{-m_{\omega }\,r}}{r}  \label{yukasolo}
\end{equation}

At the first order in the momentum transfer, in the present configuration, $%
\Phi (x)$ must include only the central and spin-spin components of the
nuclear potential. Indeed, it can be easily seen that the spin-orbit and
tensor components vanish in this ground state configuration \cite{cohen}. We
use the expressions of the Bonn potential \cite{mach} coming from the
exchange of $\sigma $ and $\omega $ mesons and including the monopolar form
factors in order to account phenomenologically for the extended hadronic
structure. Then

\begin{eqnarray}
\Phi (x) &=&-\frac{g_{\sigma h}^{2}}{4\pi }\left( 1-\frac{1}{4}\left( \frac{%
m_{\sigma }}{M_{h}}\right) ^{2}\right) \left[ \frac{e^{-m_{\sigma }r}}{r}-%
\frac{e^{-\Lambda _{\sigma }r}}{r}-\frac{(\Lambda _{\sigma }^{2}-m_{\sigma
}^{2})}{2\Lambda _{\sigma }}e^{-\Lambda _{\sigma }r}\right]  \nonumber \\
&&+\frac{g_{\omega h}^{2}}{4\pi }\left( 1+\frac{1}{3}\left( \frac{m_{\omega }%
}{M_{h}}\right) ^{2}\right) \left[ \frac{e^{-m_{\omega }\,r}}{r}-\frac{%
e^{-\Lambda _{\omega }\,r}}{r}-\frac{(\Lambda _{\omega }^{2}-m_{\omega }^{2})%
}{2\Lambda _{\omega }}e^{-\Lambda _{\omega }\,r}\right]  \label{yukacss}
\end{eqnarray}
is the half of the interaction energy between two antiparallel-spin hyperons
at a point and an point-like hyperon at a distance $x$, including the
central and spin-spin components as well as the form factors.

Now, the total potential energy of an hyperon around a lattice site is
obtained as a superposition of the interaction potential energies with the
hyperons centered on the other sites,

\begin{equation}
U_{\sup }(\overrightarrow{r})=\sum_{i}U(|\overrightarrow{r}-\overrightarrow{r%
}_{i}|)
\end{equation}
\label{eq:(2.8)}to which the interaction energy with the fluid of nucleons
must be added. For a cubic lattice, the positions of the sites are defined as

\[
\vec{r}_{i}=a(l\vec{i}+m\vec{j}+n\vec{k}) 
\]
\label{eq:(2.9)}$l$, $m$ and $n$ being null, positive or negative integers
and $a$ is the lattice constant which is related to the mean hyperonic
density by $<n_{h}>_{s}=2/a^{3}$ (the index ''s'' stands for ''spatial
average). The case where the three indexes vanish simultaneously is
explicitly excluded and, in this way, the interaction energy between the two
hyperons at each site (as well as the self interaction) is excluded. As
mentioned in the introduction, the contribution of this energy to the total
energy of the system, as estimated from the present approximation, is small,
and will be treated as a perturbation, which will be incorporated in the
final calculation of the thermodynamic state.

As we shall see later, for many reasonable choices of the model parameters,
this potential has the form of a confining nearly parabolic well at every
lattice site. We can describe the dynamics of the hyperon in this well by
solving the Schr\"{o}dinger equation, and in this way determine
self-consistently the oscillator frequency which enters the expression of
the potential energy in (\ref{pot}). Indeed, the frequency of the harmonic
oscillator is related to the second derivative of the harmonic potential by

\begin{equation}
\nu _{0}=\sqrt{\frac{(\nabla ^{2}U(\overrightarrow{r}))_{\overrightarrow{r}%
=0}}{M_{h}}}  \label{frec}
\end{equation}
\label{eq:(2.10)}where U is the parabolic fit of $U_{sup}$ around a site.

The nucleonic sector is described here as a uniform Fermi fluid in the mean
field approximation \cite{wal}, but undergoing the action of the spatial
average of the lattice fields, besides the one of the mean fields generated
by the nucleon distribution itself. In this way, we calculate the
contribution of the nucleons to the total energy density of the crystallized
system in the same approximation as the one used in the Fermi liquid case 
\cite{glenlibro}, to which we will compare in order to evaluate the relative
stability of the liquid and solid configurations.

The equations for the scalar and vector fields in the medium are

\begin{equation}
(-\nabla ^{2}+m_{\sigma }^{2})(<\sigma >+\sigma _{h})=g_{\sigma
N}S_{N}+g_{\sigma h}S_{h}  \label{ecs}
\end{equation}
\begin{equation}
(\nabla ^{2}-m_{\omega }^{2})(<\omega ^{0}>+\omega _{h})=g_{\omega
N}n_{N}+g_{\omega h}n_{h}  \label{ecw}
\end{equation}
\label{eq:(2.12)}whereas the equations for the mean fields generated by the
uniform nucleon (proton and neutron) distributions are

\begin{eqnarray}
m_{\sigma }^{2}(\frac{m_{N}-M_{N}}{g_{\sigma N}}) &=&\frac{g_{\sigma N}}{%
2\pi ^{2}}M_{N}\text{ }(p_{fp}\varepsilon _{fp}-M_{N}^{2}\ln \QDOVERD| |
{p_{fp}+\varepsilon _{fp}}{M_{N}}+p_{fn}\varepsilon _{fn}-M_{N}^{2}\ln
\QDOVERD| | {p_{fn}+\varepsilon _{fn}}{M_{N}})  \nonumber \\
+g_{\sigma h} &<&n_{h}>_{s}  \label{sig}
\end{eqnarray}
\begin{equation}
m_{\omega }^{2}<\omega ^{0}>=-g_{\omega N}\left( \frac{p_{fp}^{3}}{3\pi ^{2}}%
+\frac{p_{fn}^{3}}{3\pi ^{2}}\right)  \label{omega}
\end{equation}
\label{eq:(2.12bis)}where $\varepsilon _{fp,n}=\sqrt{p_{fp,n}^{2}\ +\
M_{N}^{2}}$ and the effective mass of the quasi-nucleons is now given by

\begin{equation}
M_{N}=m-g_{\sigma }(<\sigma >+<\sigma _{h}>_{s})
\end{equation}
\label{eq:(2.13)}In the above equations $<\sigma _{h}>_{s}$ and $<n_{h}>_{s}$
are the spatial averages of the hyperonic contributions to the scalar field
and the hyperon density respectively. Such spatial averages are defined as
the limit of

\begin{equation}
<f(x)>_{s}={\frac{1}{V}}\int d_{3}\,x\ f(x)
\end{equation}
\label{eq:(2.14)}as the volume $V$ goes to infinity. The spatial averages of
the lattice scalar and vector fields are related to the average hyperon
density through 
\begin{eqnarray}
m_{\sigma }^{2} &<&\sigma _{h}>_{s}=g_{\sigma h}<S_{h}>_{s} \\
m_{\omega }^{2} &<&\omega _{h}>_{s}=-g_{\omega h}<n_{h}>_{s}
\end{eqnarray}
\label{eq:(2.15)}In the non-relativistic approximation for the hyperons in
the lattice, the scalar density and the density of hyperons coincide.

\section{The $\beta$ equilibrium}

The above description allows the complete determination of the macroscopic
configuration of the system at every fixed baryonic density. Indeed, we can
now easily calculate the chemical potentials for each component and
establish the set of equations for the beta equilibrium in the medium. The
proton, neutron and electron chemical potentials in terms of their Fermi
momentums are given by

\begin{eqnarray}
\mu _{e} &=&\sqrt{p_{fe}^{2}+m_{e}^{2}}=\varepsilon _{fe} \\
\mu _{n} &=&\sqrt{p_{fn}^{2}+M_{N}^{2}}-g_{\omega N}(<\omega ^{0}>+<\omega
_{h}>_{s}) \\
\mu _{p} &=&\sqrt{p_{fp}^{2}+M_{N}^{2}}-g_{\omega N}(<\omega ^{0}>+<\omega
_{h}>_{s})
\end{eqnarray}
\label{eq:(3.1)}The hyperon chemical potential in the ordered phase is the
energy gained by the system when adding a new hyperon in the lowest energy
empty state, which in the postulated configuration is the energy of the
first unoccupied level of the harmonic oscillator (including now the mass
and well energies) plus the interaction energy of the hyperon with the mean
fields created by the nucleon component:

\begin{equation}
\mu _{h}=M_{h}-|Uo|+\frac{5}{2}\nu _{0}-g_{\omega h}<\omega ^{0}>
\end{equation}
\label{eq:(3.2)}where the effective hyperon mass is determined by the mean
scalar field generated by the nucleon distribution through

\begin{equation}
M_{h}=m_{h}-g_{\sigma h}<\sigma >  \label{mh}
\end{equation}
\label{eq:(3.2)a}The above considerations lead to the following system of
equations for the beta equilibrium

\begin{equation}
n_{e}=n_{p}
\end{equation}
\label{eq:(3.3)}from charge neutrality,

\begin{equation}
\mu _{n}=\mu _{p}+\mu _{e}
\end{equation}
\label{eq:(3.4)}from neutron decay,

\begin{equation}
\mu _{n}=\mu _{h}
\end{equation}
\label{eq:(3.5)}from hyperon decay,

\begin{equation}
n_{b}=n_{n}+n_{p}+n_{h}
\end{equation}
\label{eq:(3.6)}from baryonic number conservation.

Moreover, the crystal cell size, $a$, is related to the macroscopic density
of hyperons, $n_{h}$, paired in S-states on the nodes, through

\begin{equation}
n_{h}=\frac{2}{a^{3}}
\end{equation}
\label{eq:(3.7)}These equations, coupled to the mean field equations (\ref
{sig}, \ref{omega}) and the self-consistent equations for the confining
fields in the lattice nodes (\ref{frec}), form a system whose solution (when
found, for a given set of the parameters of the model) completely determine
the postulated configuration for every baryonic density. The relative
stability of such ordered configuration with respect to the Fermi liquid
configuration at the same baryonic density must be established by comparing
the energy densities in both cases.

For the Fermi liquid configuration we can establish on similar grounds a
system of equations for the $\beta $ equilibrium \cite{glenlibro}. In this
case the hyperon chemical potential is given in function of the Fermi energy
by

\begin{equation}
\mu _{h}=\sqrt{p_{fh}^{2}+M_{h}^{2}}-g_{\omega h}<\omega ^{0}>
\label{eq:(3.8)}
\end{equation}
where the effective hyperon mass is now given by the same equation (\ref{mh}%
) in terms of the (now unique) mean scalar field which results from the Eq. (%
\ref{sigliq}) below. The system of equations defining the equilibrium is now:

\begin{eqnarray}
m_{\sigma }^{2}<\sigma > &=&\frac{g_{\sigma N}}{2\pi ^{2}}%
M_{N}[p_{fp}\varepsilon _{fp}-M_{N}^{2}\ln \QDOVERD| | {p_{fp}+\varepsilon
_{fp}}{M_{N}}+p_{fn}\varepsilon _{fn}-M_{N}^{2}\ln \QDOVERD| |
{p_{fn}+\varepsilon _{fn}}{M_{N}}]  \label{sigliq} \\
&&+\frac{g_{\sigma h}}{2\pi ^{2}}M_{h}[p_{fh}\varepsilon _{fh}-M_{h}^{2}\ln
\QDOVERD| | {p_{fh}+\varepsilon _{fh}}{M_{h}}]  \nonumber
\end{eqnarray}
\label{eq:(3.9)}for the mean scalar field,

\begin{equation}
m_{\omega }^{2}<\omega ^{0}>=-g_{\omega N}(\frac{p_{fp}^{3}}{3\pi ^{2}}+%
\frac{p_{fn}^{3}}{3\pi ^{2}})-g_{\omega h}\frac{p_{fh}^{3}}{3\pi ^{2}}
\end{equation}
\label{eq:(3.10)}for the mean vector field,

\begin{equation}
p_{fp}=p_{fe}
\end{equation}
\label{eq:(3.11)}from charge conservation,

\begin{equation}
\varepsilon _{fn}-g_{\omega N}<\omega ^{0}>=\varepsilon _{fp}-g_{\omega
N}<\omega ^{0}>+\sqrt{p_{fe}^{2}+m_{e}^{2}}
\end{equation}
\label{eq:(3.12)}from neutron decay,

\begin{equation}
\varepsilon _{fn}-g_{\omega N}<\omega ^{0}>=\sqrt{p_{fh}^{2}+M_{h}^{2}}%
-g_{\omega h}<\omega ^{0}>  \label{eq:(3.13)}
\end{equation}
from hyperon decay

\begin{equation}
n_{b}=\frac{p_{fp}^{3}}{3\pi ^{2}}+\frac{p_{fn}^{3}}{3\pi ^{2}}+\frac{%
p_{fh}^{3}}{3\pi ^{2}}  \label{eq:(3.14)}
\end{equation}
from baryonic number conservation.

The energy density in both cases is obtained as the sum of the contributions
of the nucleonic and hyperonic sectors and that of the fields. The spatially
averaged contribution of the fields to the total energy density in the
ordered phase in the present approximation is

\[
<\rho _{fields}>_{s}=\frac{1}{2}(m_{\sigma }^{2}<(<\sigma >+\sigma
_{h})^{2}>_{s}-\frac{1}{2}m_{\omega }^{2}<(<\omega ^{0}>+\omega
_{h})^{2}>_{s} 
\]
\begin{equation}
+\frac{1}{2}<\vec{\nabla}(\sigma _{h}).\vec{\nabla}(\sigma _{h})>_{s}-\frac{1%
}{2}<\vec{\nabla}(\omega _{h}).\vec{\nabla}(\omega _{h})>_{s}
\end{equation}
\label{eq:(3.15)}In averaging the terms containing the spatial derivatives
of the lattice fields we obtain, by integration by parts 
\begin{equation}
{\frac{1}{V}}\int \vec{\nabla}(\sigma _{h}).\vec{\nabla}(\sigma _{h})d_{3}%
\overrightarrow{x}={\frac{1}{V}}\int \vec{\nabla}.[\sigma _{h}.\vec{\nabla}%
(\sigma _{h})]d_{3}\overrightarrow{x}-{\frac{1}{V}}\int [\sigma _{h}\,\nabla
^{2}(\sigma _{h})]d_{3}\overrightarrow{x}
\end{equation}
and 
\[
{\frac{1}{V}}\int \vec{\nabla}(\omega _{h}).\vec{\nabla}(\omega _{h})d_{3}%
\overrightarrow{x}={\frac{1}{V}}\int \vec{\nabla}.[\omega _{h}.\vec{\nabla}%
(\omega _{h})]d_{3}\overrightarrow{x}-{\frac{1}{V}}\int [\omega _{h}\,\nabla
^{2}(\omega _{h})]d_{3}\overrightarrow{x} 
\]
In the limit of large V, the terms associated to the integrals of
divergences in these formulae vanish. Using the field equations (\ref{ecs}, 
\ref{ecw}) for the Laplace operator we obtain 
\begin{equation}
{\frac{1}{V}}\int \vec{\nabla}(\sigma _{h}).\vec{\nabla}(\sigma _{h})d_{3}%
\overrightarrow{x}=-{\frac{1}{V}}\int [m_{\sigma }^{2}.\sigma
_{h}^{2}+g_{\sigma h}\,\sigma _{h}\,S_{h}]d_{3}\overrightarrow{x}
\end{equation}
\begin{equation}
{\frac{1}{V}}\int \vec{\nabla}(\omega _{h}).\vec{\nabla}(\omega _{h})d_{3}%
\overrightarrow{x}={\frac{1}{V}}\int [m_{\omega }^{2}.\omega
_{h}^{2}-g_{\omega h}\,\omega _{h}\,n_{h}]d_{3}\overrightarrow{x}
\end{equation}
and finally, the expression for the spatially averaged field energy is

\[
<\rho _{fields}>_{s}=\frac{g_{\sigma N}}{2}(<\sigma >+<\sigma
_{h}>_{s})S_{N}+\frac{g_{\omega N}}{2}(<\omega ^{0}>+<\omega _{h}>_{s})n_{N} 
\]
\begin{equation}
+\frac{g_{\sigma h}}{2}<\sigma ><S_{h}>_{s}+\frac{g_{\omega h}}{2}<\omega
><n_{h}>_{s}+\frac{g_{\sigma h}}{2}<\sigma _{h}S_{h}>_{s}+\frac{g_{\omega h}%
}{2}<\omega _{h}n_{h}>_{s}  \label{densen}
\end{equation}
\label{eq:(3.20)}For the Fermi liquid configuration the field contribution
to the energy density is

\begin{equation}
\rho _{fields}=\frac{1}{2}m_{\sigma }^{2}<\sigma >^{2}-\frac{1}{2}m_{\omega
}^{2}<\omega ^{0}>^{2}  \label{campos}
\end{equation}
The contribution to the energy density of the nucleonic Fermi fluid sector
and the gas of electrons for the ordered configuration is 
\begin{eqnarray}
\rho _{p}+\rho _{n}+\rho _{e} &=&\frac{1}{8\pi ^{2}}[p_{fp}\varepsilon
_{fp}(p_{fp}^{2}+\varepsilon _{fp}^{2})-M_{N}^{4}\ln \QDOVERD| |
{p_{fp}+\varepsilon _{fp}}{M_{N}}  \nonumber \\
&&+p_{fn}\varepsilon _{fn}(p_{fn}^{2}+\varepsilon _{fn}^{2})-M_{N}^{4}\ln
\QDOVERD| | {p_{fn}+\varepsilon _{fn}}{M_{N}}+\frac{p_{fe}^{4}}{8\pi ^{2}}
\label{rho}
\end{eqnarray}
\label{eq:(3.22)}whereas the spatially averaged contribution of the
hyperonic sector in this case is given by the sum of the energies of the
oscillators in their ground states 
\begin{equation}
\rho _{h}=(M_{h}+\frac{3}{2}\nu _{0}-g_{\omega h}<\omega ^{0}>-|U_{0}|)n_{h}
\label{rhoh}
\end{equation}
\label{eq:(3.23)}where the proper energy of the quasi-hyperons has been
included in order to recover the (approximate) relativistic expression,
consistent with the treatment of the other components.

For the Fermi liquid configuration the contribution to the energy density of
the nucleons, hyperons and electrons is given by the same expression (\ref
{rho}) where the sum must now be extended to the three sectors.

Note that in both configurations the field energy density compensates the
double counting of the interaction energy density when calculated as the sum
of the energies of the individual particles. The divergent contributions of
the self energy of the hyperons in the ordered configuration contained in
the two last terms of the field energy (\ref{densen}) must be discarded.

\section{Energetic balance and stable configurations}

The existence of a stable solution for the ordered phase is sensitive to the
choice of the parameters ot the model, some of which (in particular the
strengths of the hyperonic couplings) are poorly known. Moreover, the
nuclear interaction is treated at an elementary level and in this simplified
version of the model. Consequently, no definitive conclusions about the
presence of such ordered states in actual hadronic matter can be inferred
from our analysis. Nevertheless, we shall explore their compatibility with
our knowledge of the hadronic interactions at the level of description used
in the present framework.

The values of the masses and meson-nucleon coupling constants used in the
following calculations are 
\begin{eqnarray}
m_{N} &=&939\text{ }MeV,\text{ }m_{\sigma }=550\text{ }MeV,\text{ }m_{\omega
}=783\text{ }MeV  \nonumber \\
m_{h} &=&m_{\Lambda }=1115\text{ }MeV,\text{ }\dfrac{g_{\sigma N}^{2}}{4\pi }%
=7.29,\text{ }\dfrac{g_{\omega N}^{2}}{4\pi }=10.83  \label{parametros}
\end{eqnarray}
\label{eq:(4.1)}With these values, saturation for nuclear matter is
attained, in the mean field approximation, at $p_{f0}=1.42$ $fm^{-1}$
corresponding to a particle density $n_{0}=0.197$ $fm^{-3}$, with a binding
energy $E_{b}=-15.77$ $MeV.$ The meson-hyperon coupling constants in-medium,
giving the interaction with the mean fields, are determined by fixing the
ratios

\begin{eqnarray}
x_{\sigma } &=&\frac{g_{\sigma h}}{g_{\sigma N}} \\
x_{\omega } &=&\frac{g_{\omega h}}{g_{\omega N}}
\end{eqnarray}
\label{eq:(4.2)} in such a way that the correct value of the $\Lambda $
binding energy in hypernuclei will be reproduced. Figure 1 is a diagram $%
x_{\sigma }-x_{\omega }$ showing the region of values of these parameters
which allows the fit of the binding energy of the hyperons, extracted from
spectroscopic data on single $\Lambda -$ hypernuclei \cite{Glendenning},
once the values of the meson-nucleon coupling constants at saturation
density are fixed by the parametrization (\ref{parametros}).

On the other hand, for the ordered configuration, the coupling of the
hyperons to the scalar and vector fields governing the interaction in the
lattice, which has been introduced as an interaction in vacuum (see Eq. (\ref
{v}) ) will be fixed to different values. In reference \cite{reuber94} the
coupling of the  $\Lambda $ hyperons to the $\sigma $ field and the vector
coupling to the $\omega $ field are taken to be $\dfrac{g_{\sigma h}}{\sqrt{%
4\pi }}=2.138$, $\dfrac{g_{\omega h}}{\sqrt{4\pi }}=2.981$ in order to fit
the scattering data in vacuum. Although our level of description of the
interaction is more elementary than in the above mentioned reference, we
shall keep these values in our calculations as a first approach.

Let us first consider a solution of the system for the ordered phase in the
simplest case where the form factors are discarded and the first order
corrections to the hyperon potential in the momentum transfer are neglected
(using Eq. (\ref{yukasolo})). The parameters $x_{\sigma }$ and $x_{\omega }$
are now fixed to the values $0.45$ and $0.483$ respectively (set A), which
fit the $\Lambda -$ hypernuclei binding energy. Simultaneously, the system
for the liquid configuration is solved with the same set of parameters. In
figure 2 (curve A) we show the difference between the energy densities in
the liquid and the ordered configurations as a function of the baryonic
density. Before the threshold density for the creation of hyperons ($%
n_{t}=1.75$ $n_{0}$) the liquid phase of protons, neutrons and electrons is
the only stable configuration. Beyond this threshold the solid configuration
is present and more stable than the liquid one. This relative stability
increases with density.

When the first order corrections in the momentum transfer are included in
the hyperonic potential (\ref{yukacss}), but excluding the terms coming from
the presence of the form factors, stable ordered configurations are also
obtained for some sets of parameters. The curve B in figure 2 results from
the set $x_{\sigma }=0.5;$ $x_{\omega }=0.546$ (set B), which also fits the $%
\Lambda $ binding energy in hypernuclei. Now the solid phase is less stable
than the liquid one from the threshold density of appearance of the
hyperonic sector ($n_{t}=1.79$ $n_{0}$) up to a density of $4.3$ $n_{0}$,
around which a transition between the two phases takes place. When the form
factors are included by using the full Eq. (\ref{yukacss}) in the
calculation, stable ordered configurations of the ground state are obtained
when we fix the form factors $\Lambda _{\sigma }=1.5$ GeV and $\Lambda
_{\omega }$=2.0 GeV, and the ratios between the coupling constants to $%
x_{\sigma }=0.4,$ $x_{\omega }=0.42$ (set C), which fit also the $\Lambda $
binding energy in hypernuclei. Now, the liquid phase is more stable in a
narrow range of densities between the threshold ($n_{t}=1.72$ $n_{0}$) and $%
n\sim 2.2n_{0}$ (see Fig. 2).

Both in the B and C cases, a Gibbs construction should be performed in order
to determine the transition region \cite{glen92}. As can be seen from Fig.
2, the density around which the liquid-solid phase transition takes place is
very sensitive to the level of description of the nuclear interaction (the
optimum level in the present calculations corresponds to set C), but the
ordered phase is always more stable than the liquid one at high enough
densities.

For the solid configurations we obtain the size of the lattice cell (the
lattice constant, a) as a function of the baryonic density (see figure 3).
This parameter decreases monotonically with density (from infinity at the
threshold) in all the meaningful range. This is related to the increasing
abundance of the hyperons with baryonic density. The solution of the $\beta $%
-equilibrium system gives the abundances of all the species in each
configuration, which are plotted in figures 4(A,B,C) as functions of
density, for the three sets of parameters, and compared to the abundances in
the corresponding liquid configurations. Before the threshold densities the
hyperonic sector is absent and the neutron abundance decreases with density
with a simultaneous increase of the proton and electron abundances. Beyond
the threshold the regular increase of the abundance of hyperons in the solid
and liquid phases is accompanied by a simultaneous reduction of the
abundances of all the other fermions. Qualitatively, the behavior of the
abundances for the three sets of parameters in the solid and liquid
configurations is similar.

Figure 5 shows the typical shape of the confining potential around a lattice
site resulting from the interaction with the hyperons located in the other
sites, which is obtained selfconsistently from Eqs.(\ref{yukasolo}, \ref
{yukacss}) when solving the system of equilibrium equations. The curve
(corresponding to the set A at a density $n=3.5$ $n_{0}$) gives the
variation of the potential energy of a point-like hyperon with the distance
to the node in the direction of a principal axis of the lattice. The
potential well has a nearly spherical symmetry around the site. We have
approximated this potential by a parabolic one in the calculation of the
solid configuration. The reliability of such an approximation is related to
the depth of the potential well, which is plotted in figure 6 as a function
of density, for the three sets of parameters. We see that, beyond the
threshold density, the confining energy increases regularly and the harmonic
approximation becomes more accurate at high densities.

At lower densities, near the threshold $n_{t}$, the abundance of hyperons is
small, both in the solid and liquid configurations. Thus the hyperons in the
liquid phase are not very degenerate and their dynamics is governed by
small-momentum wave functions. In the solid phase the hyperons are not
strongly confined, as can be seen in figure 7, where we have plotted the
ratio between the width of their gaussian wave functions to the lattice
parameter (which is very large near the threshold $n_{t}$), versus hadronic
density.

Consequently, the widths of the gaussian wave functions diverge at the
threshold and there is no important difference between the physical states
of the hyperons in both configurations in this region, where a very similar
macroscopic behavior of the liquid configuration and the ordered one (as
calculated from the harmonic approximation) should be expected. This is
confirmed by the small difference between the energy densities of both
configurations near the threshold for all sets of parameters, as well as by
the very similar behavior of the other macroscopic functions (see below).

Let us estimate the importance of the error introduced when we neglect the
interaction energy between the hyperons on the same site ($E_{cell}$), which
should modify the form of the confining potential. We have calculated the
ratio between $E_{cell}$ and the depth of the potential well (\TEXTsymbol{%
\vert}$U_{0}|$). This ratio is shown in figure 8 as a function of the energy
density for the three sets of parameters.

We see that near the threshold $n_{t}$, this ratio is important, but $%
E_{cell}$ is negative up to a density $\sim 3$ $n_{0}$. Then, by including
the contributions associated to $E_{cell}$ we should improve the confinement
of hyperons and the stability of the solid phase in this region.
Nevertheless, the modification of the macroscopic functions associated to $%
E_{cell},$ is expected to be small near the threshold, owing to the small
proportion of hyperons there. For higher densities the ratio remains less
than 0.2 for sets B an C and reaches 0.4 at $n\sim 5$ $n_{0}$ for set A, but
the associated corrections on the macroscopic quantities should be more
important here, owing to the increasing abundance of hyperons with density.
In order to explore the importance of these corrections on the macroscopic
state we have added the density of interaction energy between the hyperons
on the nodes to the total energy density in Fig. 2 (dotted lines). We see
that the correction remains small up to densities $\sim 3.5$ $n_{0}$ and
becomes more important beyond for all sets.

Let us now compare the behavior of the pressure in both configurations.
Although in the ordered phases the stress is described by a tensor, the
dominant terms come from the diagonal part of this stress tensor, which in
the cubic lattice configuration is given by the pressure \cite{LANDAU},
related to the energy and particle densities through

\begin{equation}
p=n_{b}^{2}{\frac{\partial (\rho /n_{b})}{\partial n_{b}}}
\end{equation}
\label{eq:(4.3)}Figures 9(A, B, C) are plots of the pressure in the liquid
and ordered configurations as functions of the baryonic density, for the
three sets of parameters. As is easily seen, the pressure in both phases has
a very similar behavior up to densities $\sim 3.5$ $n_{0}$ for the three
sets.

This similarity continues to higher densities for the set A, whose EOS in
the solid configuration becomes slightly stiffer than the liquid one. On hte
other hand for sets B and C the EOS of the solid phases at high densities
become dramatically softer than those of the liquid ones.

At lower densities the contribution from the hyperonic sector to the
thermodynamical behavior of the system is small, as compared to the
contribution from the nucleons, which is dominant and very similar in both
phases. This is the reason for the similar behavior of pressure in both
phases.

The behavior at higher densities can be explained in terms of the
characteristics of the hyperonic interaction. There, the hyperon sector
becomes dominant and behaves as a Fermi liquid in the liquid phase, where
the main contribution to the pressure comes from the kinetic energy of the
constituent particles. In the solid phase the pressure is mainly dominated
by the total energy of the harmonic oscillators on the nodes of the lattice
and their evolution with density, which is strongly dependent on the form of
the OBE potential between the hyperons used in the calculation. In figure 10
we have plotted the frequency of the harmonic oscillators ($\nu _{0}$, in
natural units) as a function of the baryonic density, for the three sets of
parameters. This frequency is proportional to the oscillator energy, which
enters the total energy density expression (\ref{rhoh}), and is related to
the lattice potential through Eq. (\ref{frec}), which is obtained from the
OBE potentials associated to each set of parameters.

In the case of set A, the lattice potential obtained from the OBE potential (%
\ref{yukacss}) evolves with the density in such a way that the slope of the $%
\nu _{0}-n$ curve increases strongly at $n\sim 3.2$ $n_{0}$ and becomes
greater than those of sets B and C at high densities (see Fig. 10). As a
consequence, the slope of the energy density of the solid grows nearly
parallel to the liquid phase one in this density region (see Fig. 2). This
explains that the pressure in both phases, which is directly related to
these slopes, grows similarly in this case.

For sets B and C the lattice potential is obtained from the OBE potential (%
\ref{yukacss}), which includes the contributions of the spin-spin component
and the first-order momentum transfer corrections to the central component.
As a consequence of these modifications, the depth of the well of the OBE
potential is reduced for these sets and the width of the resulting lattice
potential decreases with density slower than in the case of set A, at high
densities. Then, the slopes of curves B and C in Fig. 10 (as well as the
slopes of the corresponding $\rho -n$ curves) are smaller and the pressure
is reduced with respect to the cases of set A and liquid phase.

\section{Conclusions and perspectives}

We have analyzed the conditions of formation of ordered phases in hadronic
matter, due to the confinement of the hyperonic component, which must be
present beyond a threshold density, on the nodes of a lattice. The analysis
has been performed in the framework of a minimal model of the hadronic
plasma which incorporates some basic features of the nuclear interaction,
allowing for an acceptable fit of the nuclear saturation properties, hyperon
binding energy in hypernuclei and scattering data in vacuum. In the
resolution of the model for the ordered phase we have introduced some
approximations, consistent with the mean-field approximation which leads to
the macroscopic description of the hadronic liquid phase. In the framework
of this model, we conclude that the solid phase is always energetically
favored at high densities. At lower densities the relative stability of the
liquid and solid configurations is very sensitive to the level of
description of the hyperonic interaction. At the simplest level we obtain a
solid phase which is more stable than the liquid one for all densities
beyond the threshold of production of hyperons (set A). When the first-order
corrections in the momentum transfer are included in the static hyperonic
potential (set B), we obtain a solution where the liquid configuration is
more stable than the ordered one from the threshold to a density $\sim 4.2$ $%
n_{0}$. Beyond, the solid configuration becomes more stable. When we include
in the description of the interaction the effects of the form factors,
accounting for the hyperon extended structure (set C), the solid
configuration becomes more stable from densities slightly beyond the
threshold. Both in the cases B and C there should be a first-order phase
transition region between the two configurations, to be determined through
by a Gibbs construction \cite{glen92}. The important softening of the
hadronic EOS in presence of the solid phase, in the cases B and C at high
densities, as compared to the usual liquid EOS, should introduce important
modifications in the analysis of the hydrostatic equilibrium and cooling of
neutron stars.

Nevertheless, owing to the simplicity of the model and the approximations
introduced, these results should be considered as preliminary. In order to
extract reliable conclusions as to the presence of such phases in actual
hadronic matter, the present model must be improved in many aspects:

1) A more accurate description of the nuclear interaction must be
introduced, by including other meson exchanges. In particular the coupling
to $\rho $ (and perhaps, $\pi $) mesons can be crucial in this context,
owing to their isospin asymmetry.

2) The hyperonic sector must be extended in order to include all the hyperon
states, in particular the $\Sigma ^{-}$. Similar extension should be
introduced in the leptonic sector.

The dynamic analysis can be improved in many ways.

3) The Hartree-Fock higher order approximation can be used, instead of the
mean field one, in describing the proton and neutron liquids. Exchange
effects should be also included in the calculations concerning the hyperonic
sector.

4) The interaction between the nucleonic sector and the hyperonic lattice,
which is approximated as an uniform background in the present (Sommerfeld)
approximation, can be improved by solving the Dirac equation in the periodic
field created by the lattice, by generalizing the well know Bloch method for
the analysis of the dynamics of electrons in metals \cite{AS76}.

5) Related to the last point, the screening effects due to the
redistribution of the nucleons in presence of the lattice potential have
been calculated in the RPA \cite{prep}. These effects introduce corrections
in this potential (as well as in the energetic balance of the solid phase)
and give a first simple approximation to the (more precise) calculation of
the nucleon dynamics in terms of the Bloch functions. We have not yet
completely explored the consequences of these corrections, but preliminary
results show that the effects of the screening introduced by the
redistribution of the nucleons around the lattice sites improves the
confinement of the hyperons in the nodes and the stability of the solid
phase.

6) The interaction energy between the two hyperons in the potential well at
every lattice site has been considered here as a small perturbation.
Nevertheless, we can improve the calculation if this interaction is included
from the beginning in the lattice potential.

7) Finally, other configurations with single hyperons at each node (with
given spatial distributions of their spins), and/or non-cubic lattice
structures could be more stable than the one considered here. This program
is now in progress.

The authors would like to thank S. Bonazzola, B. Carter and P. Haensel for
useful discussions. M.A.P.G. ackowledges the support of Spanish M.E.C. grant
AP97-76955547. This work is partially supported by project MCT-00-BFM-0357.

\newpage

Figure captions.

Figure 1. Values of the rations $x_{\sigma }=\frac{g_{\sigma h}}{g_{\sigma N}%
}$ and $x_{\omega }=\frac{g_{\omega h}}{g_{\omega N}}$ leading to the
experimental $\Lambda $-binding energy in hypernuclei.

Figure 2. Difference between the energy densities of the liquid and solid
phases as functions of the baryonic density (in units of the saturation
density), for the three sets of parameters.

Figure 3. Lattice parameter as a function of the baryonic density for the
three sets.

Figure 4. Abundances of the fermion components in the liquid and solid
phases as functions of the baryonic density, for the three sets of
parameters.

Figure 5. Typical shape of the lattice potential energy (in units of the
nucleon proper energy) confining the hyperons in a node as a function of the
distance to the center of the node. It is obtained as the superposition of
the potential energies in the field created by the hyperons confined on the
neighbors nodes. The figure shows a section of this potential in the
direction of a principal axis and corresponds to set A, at 3.5 times
saturation density including the neighbors to the order 5.

Figure 6. Absolute value of the depth of the wells of the lattice potential
confining the hyperons on the nodes as a function of the baryonic density,
for the three sets of parameters.

Figure 7. Ratio between the width of the gaussian wave function of an
hyperon in a lattice site and the lattice parameter as a function of the
baryonic density, for the three sets of parameters.

Figure 8. Ratio between the interaction energy of two antiparallel-spin
hyperons confined on a lattice site and the depth of the confining well as a
function of the baryonic density, for the three sets of parameters.

Figure 9. Pressure (in MeV.fm$^{-3}$) as a function of the baryonic density 
(in units of the saturation density) for the liquid (dashed curves) and
solid phases (continuous curves), for the three sets of parameters.

Figure 10. Frequency of the hyperonic oscillators (in units of the nucleon
proper energy) in the lattice nodes as a function of the baryonic density
(in units of the saturation density), for the three sets of parameters.

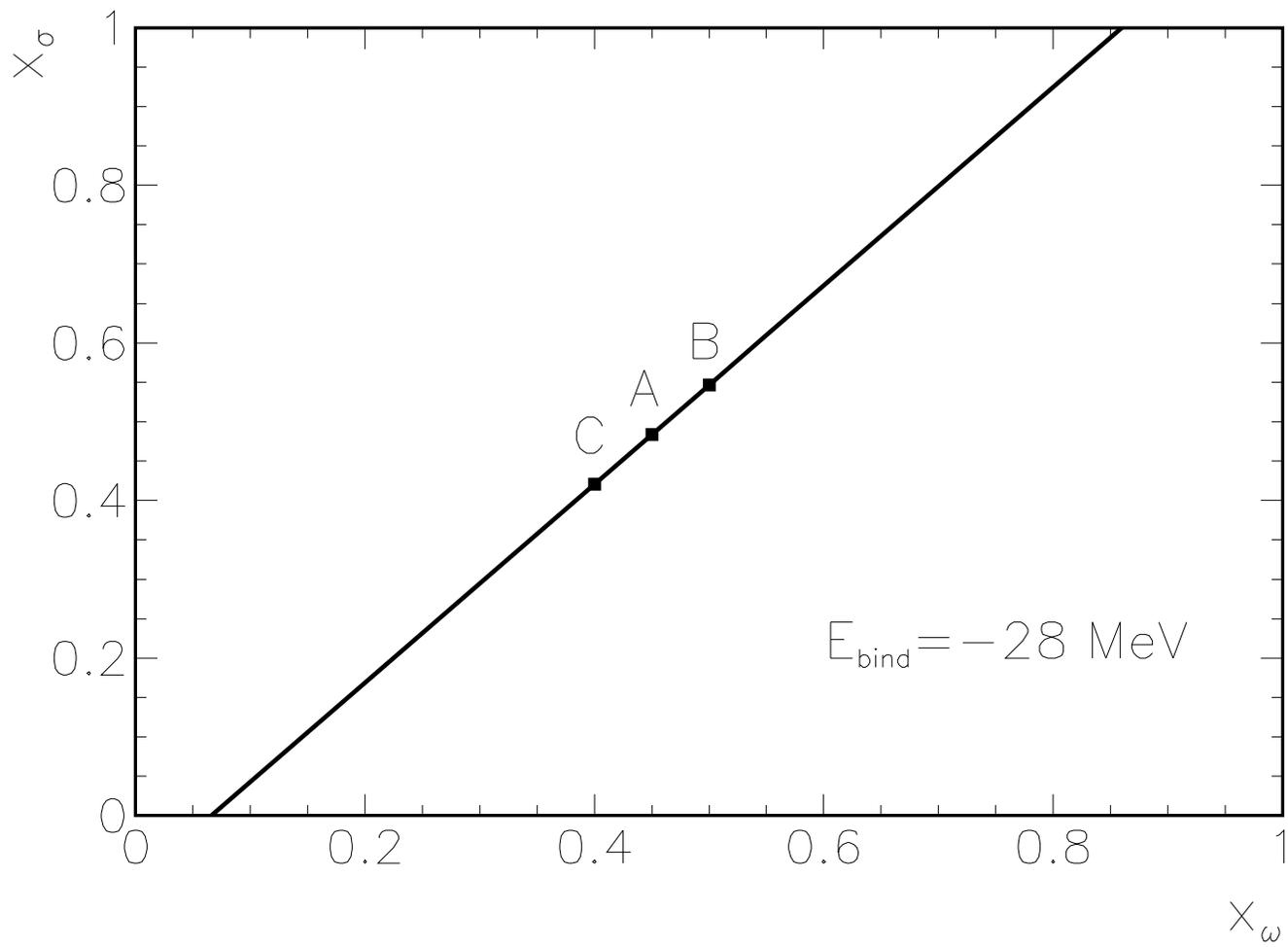
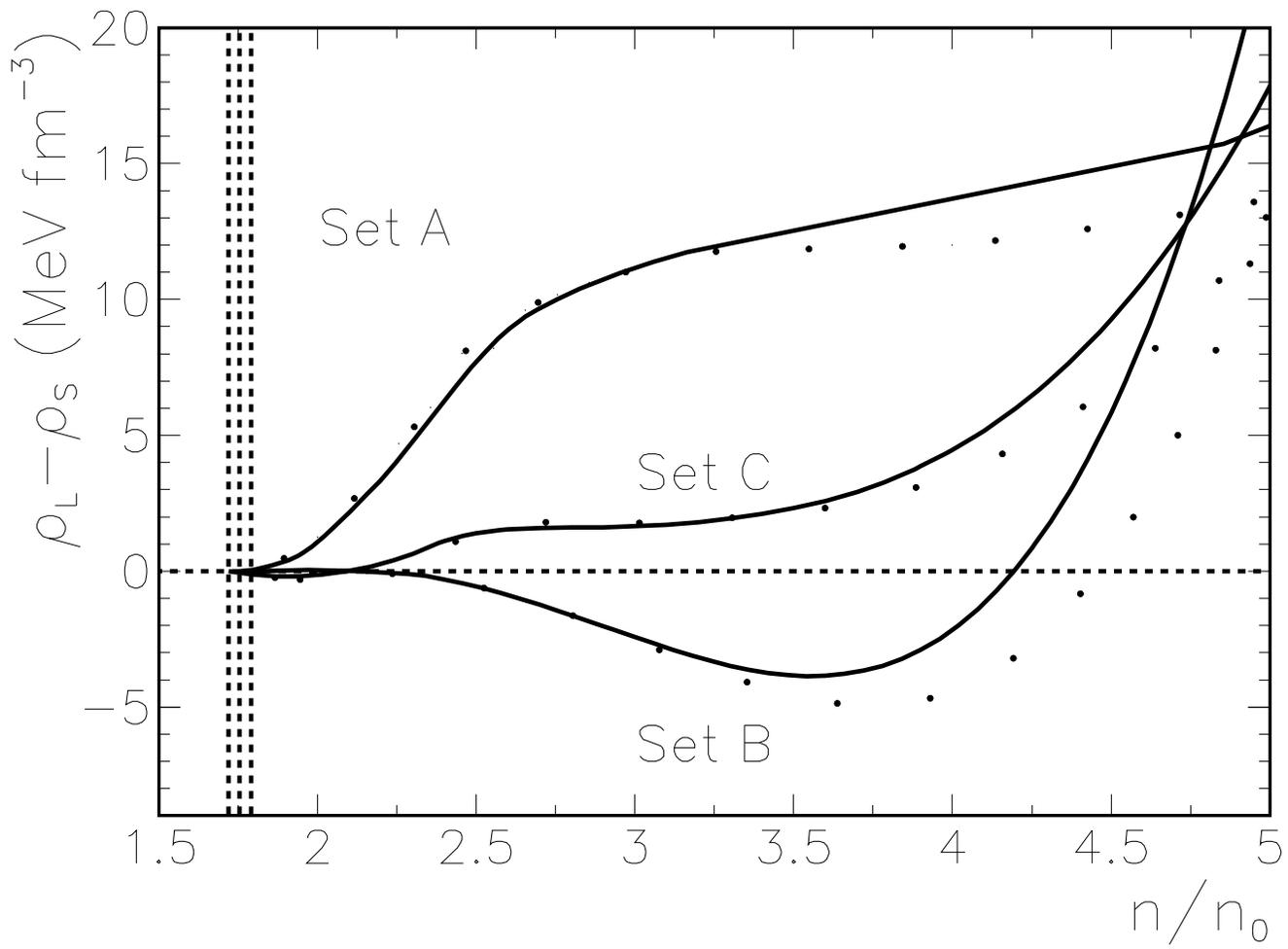
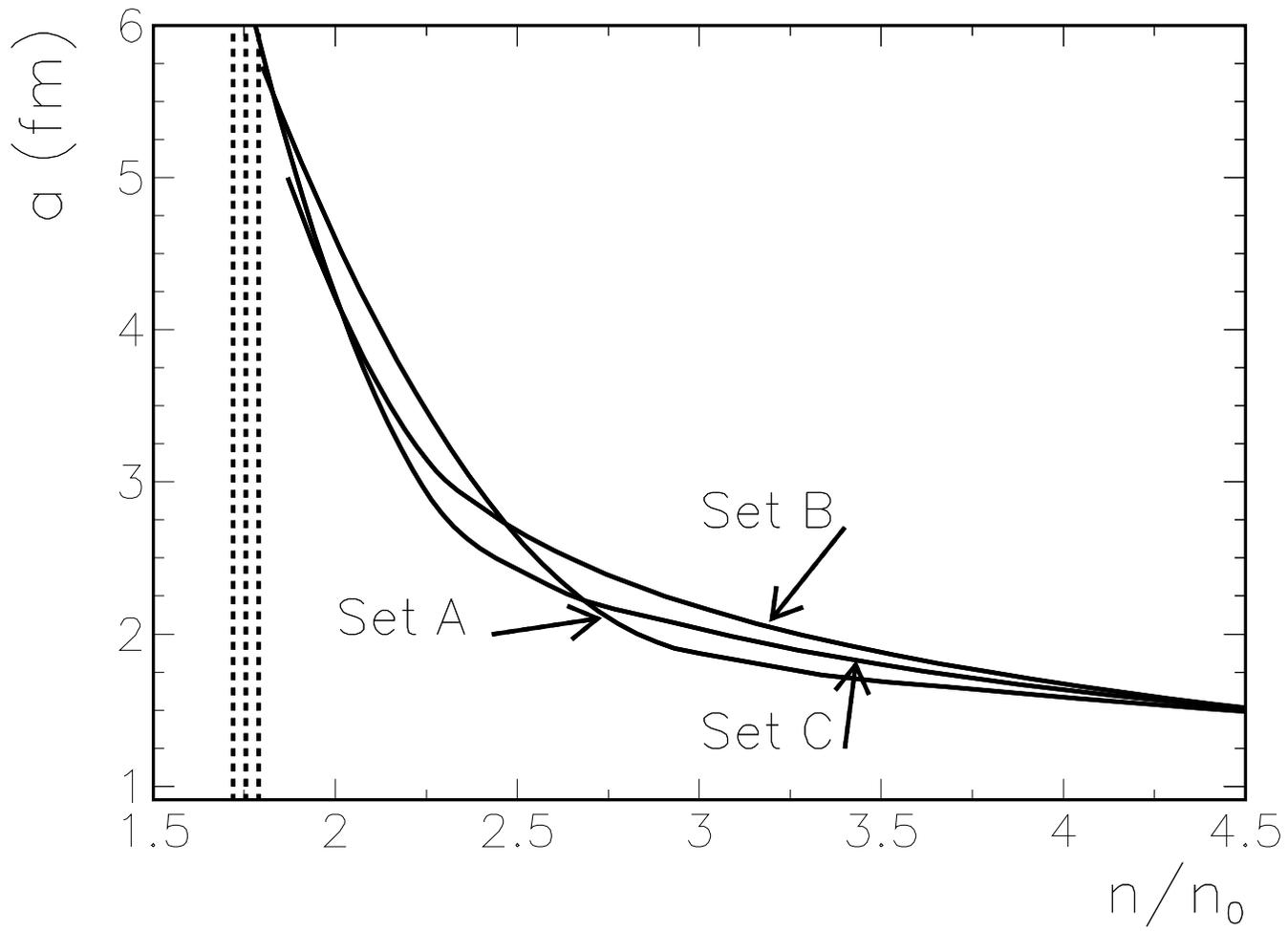
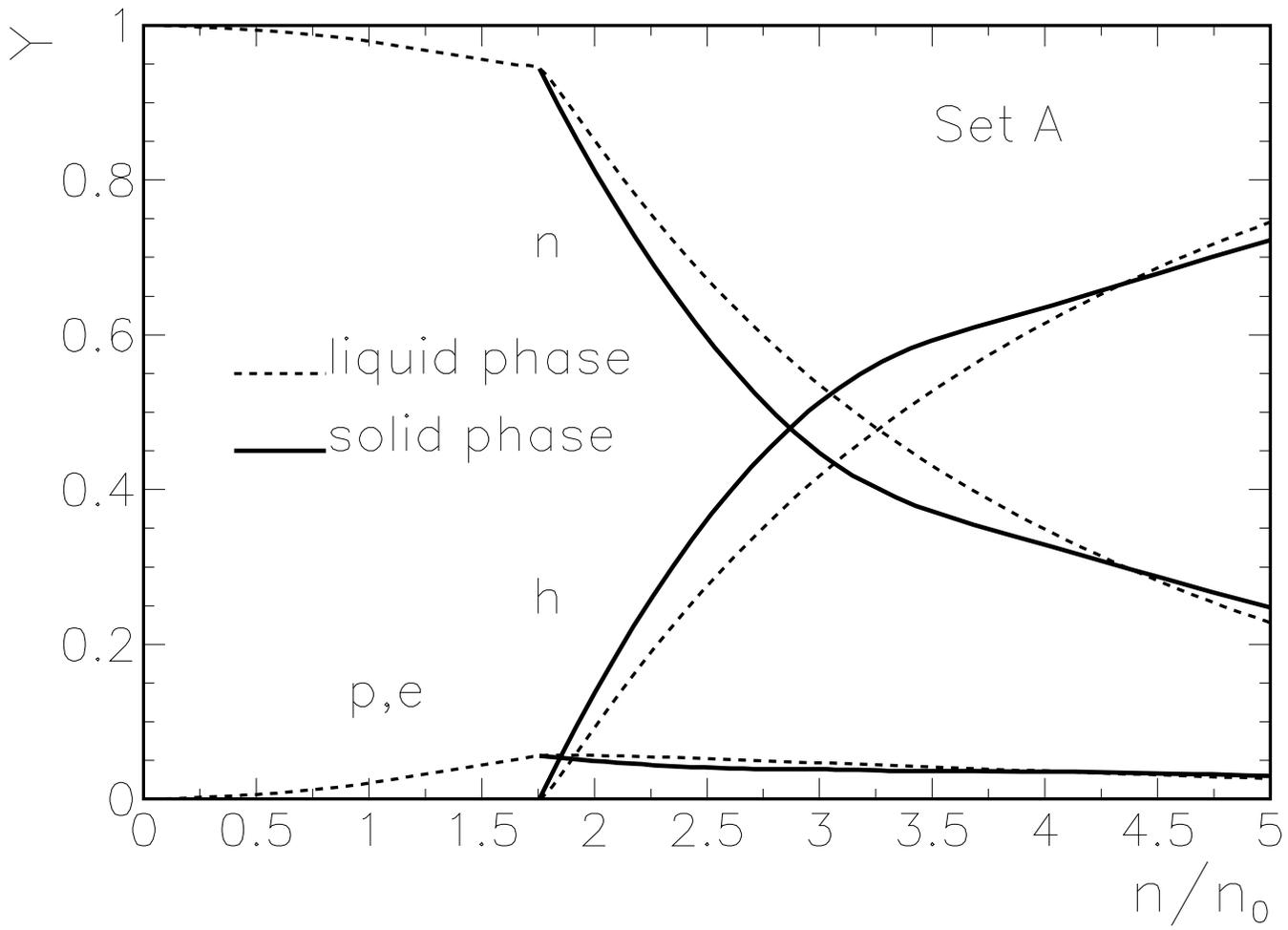
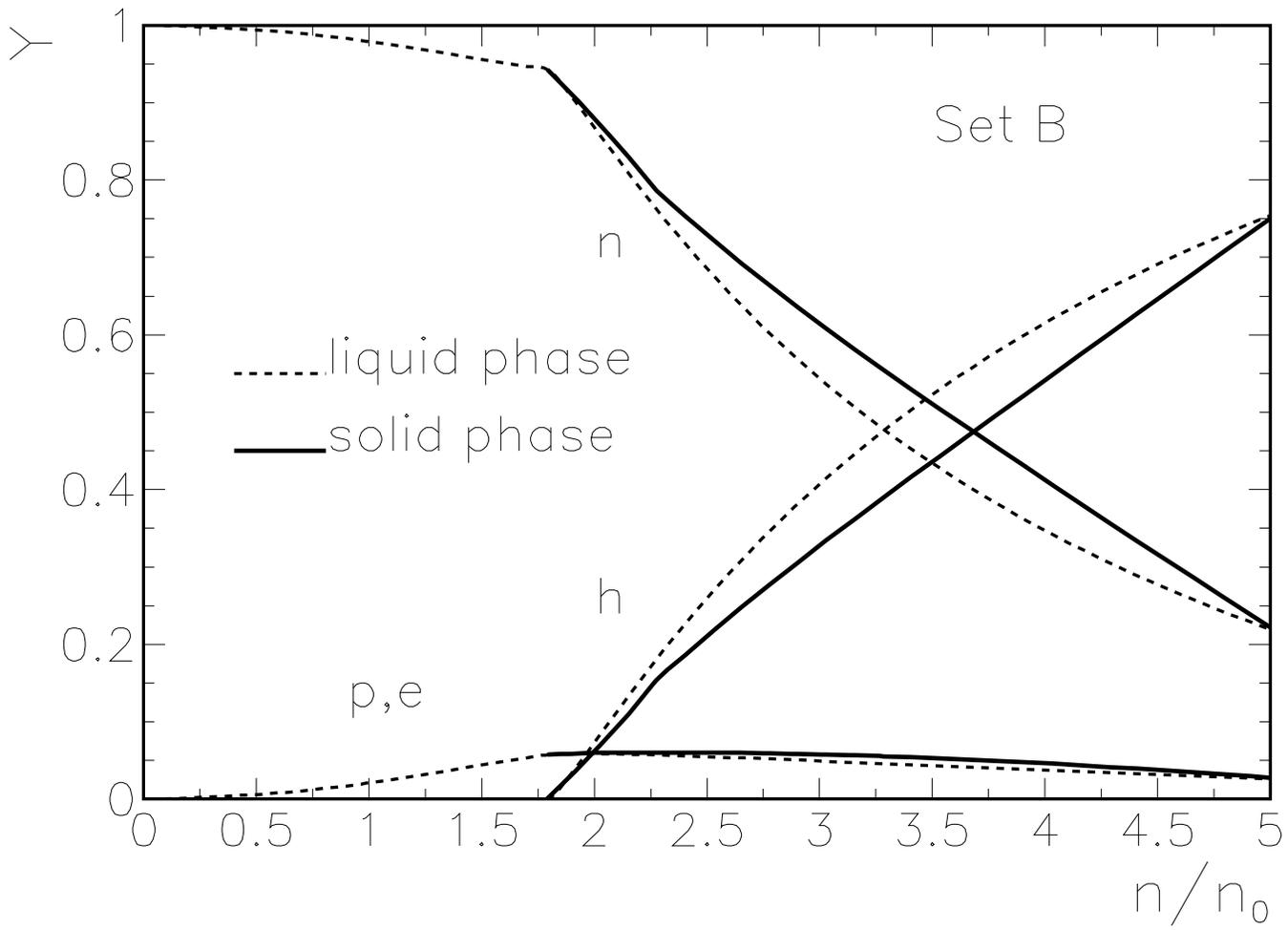
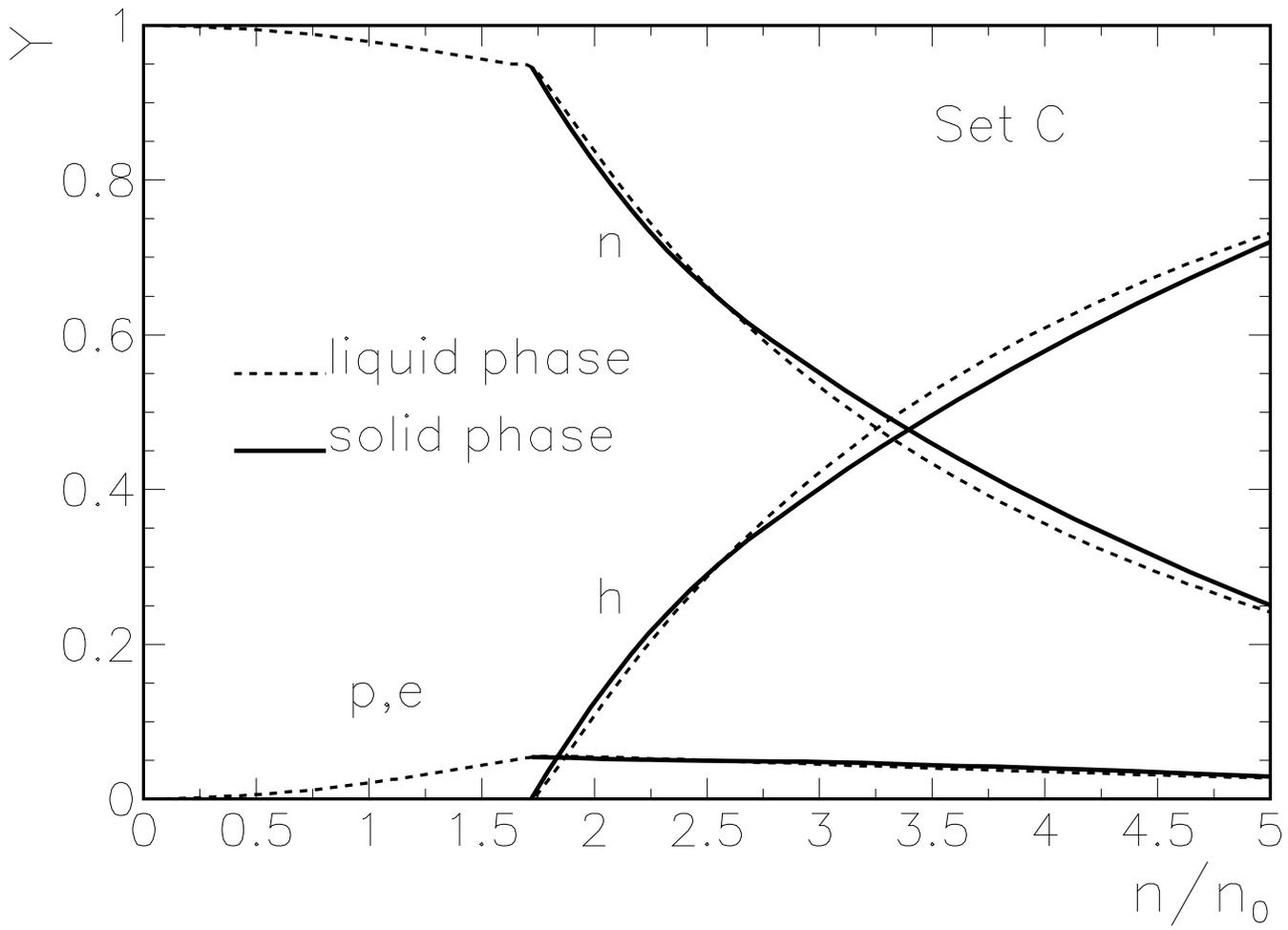
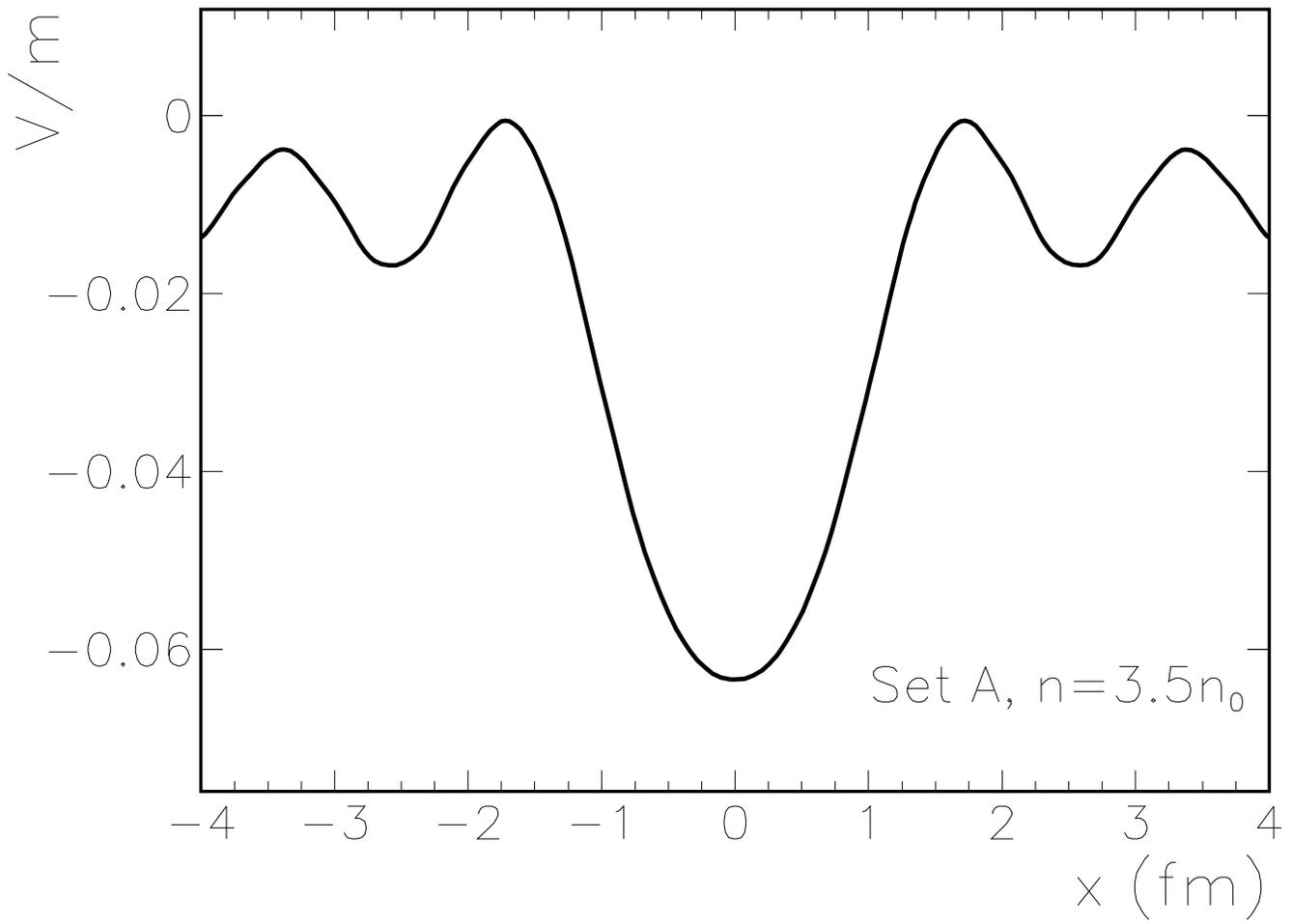
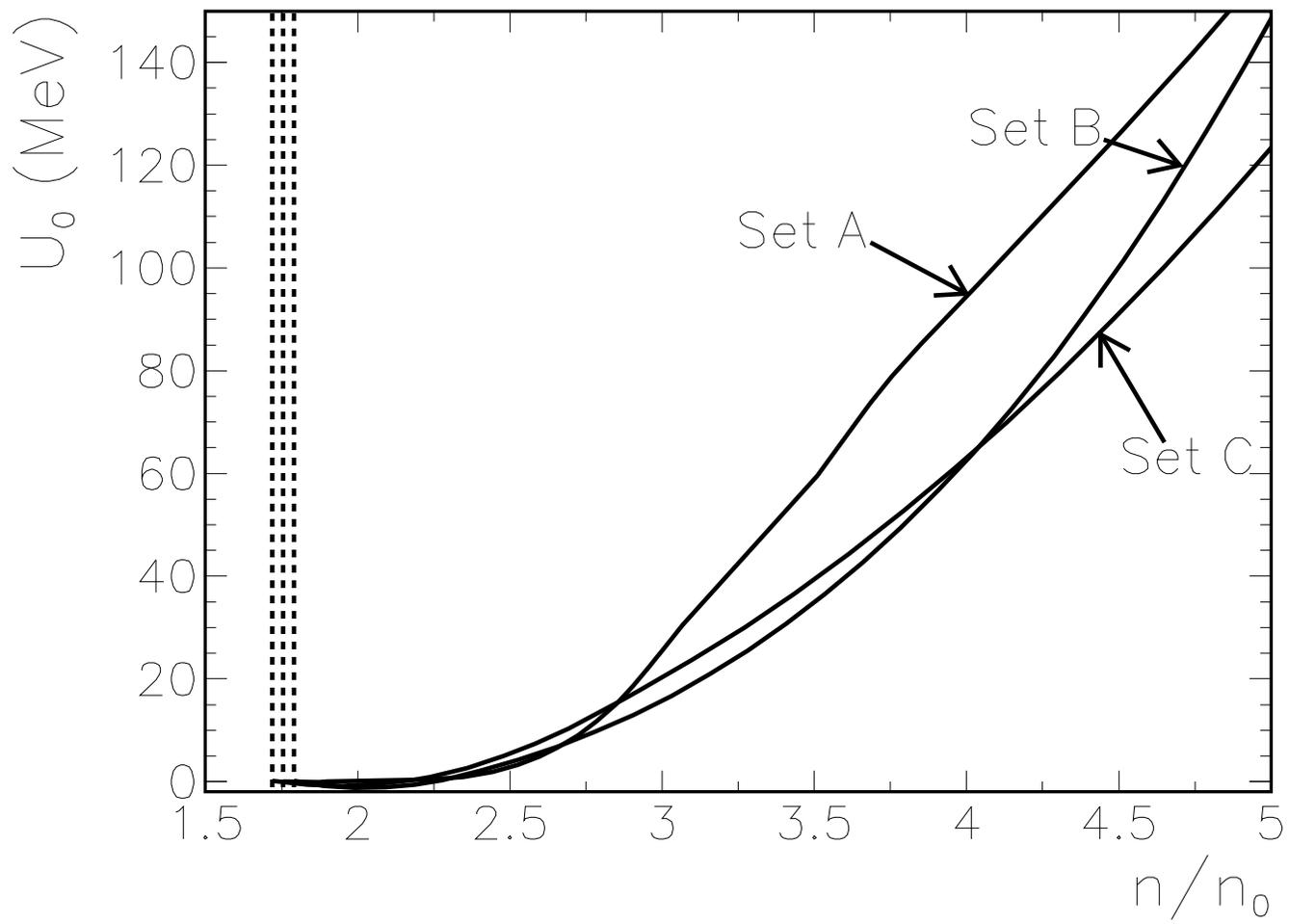
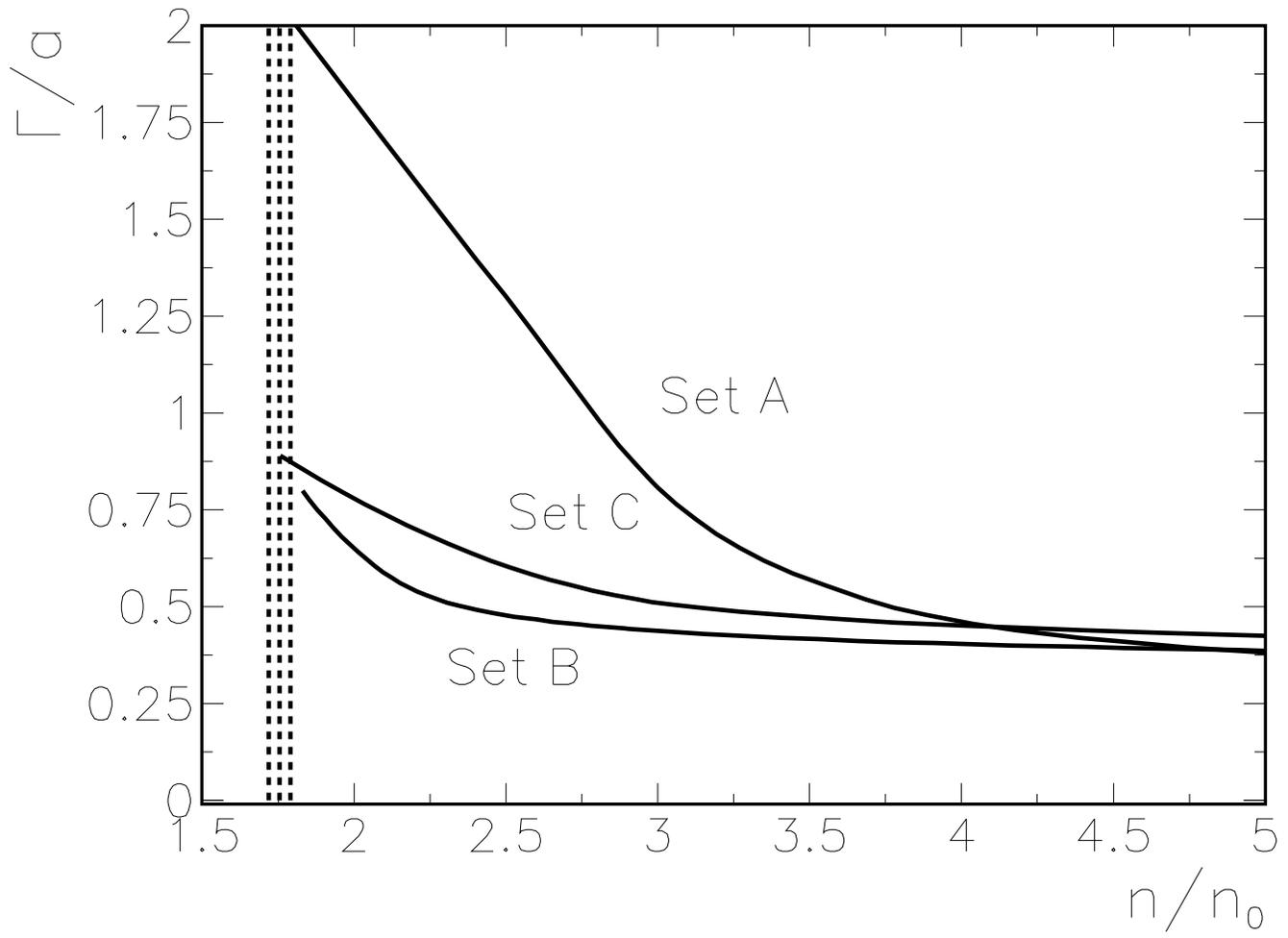
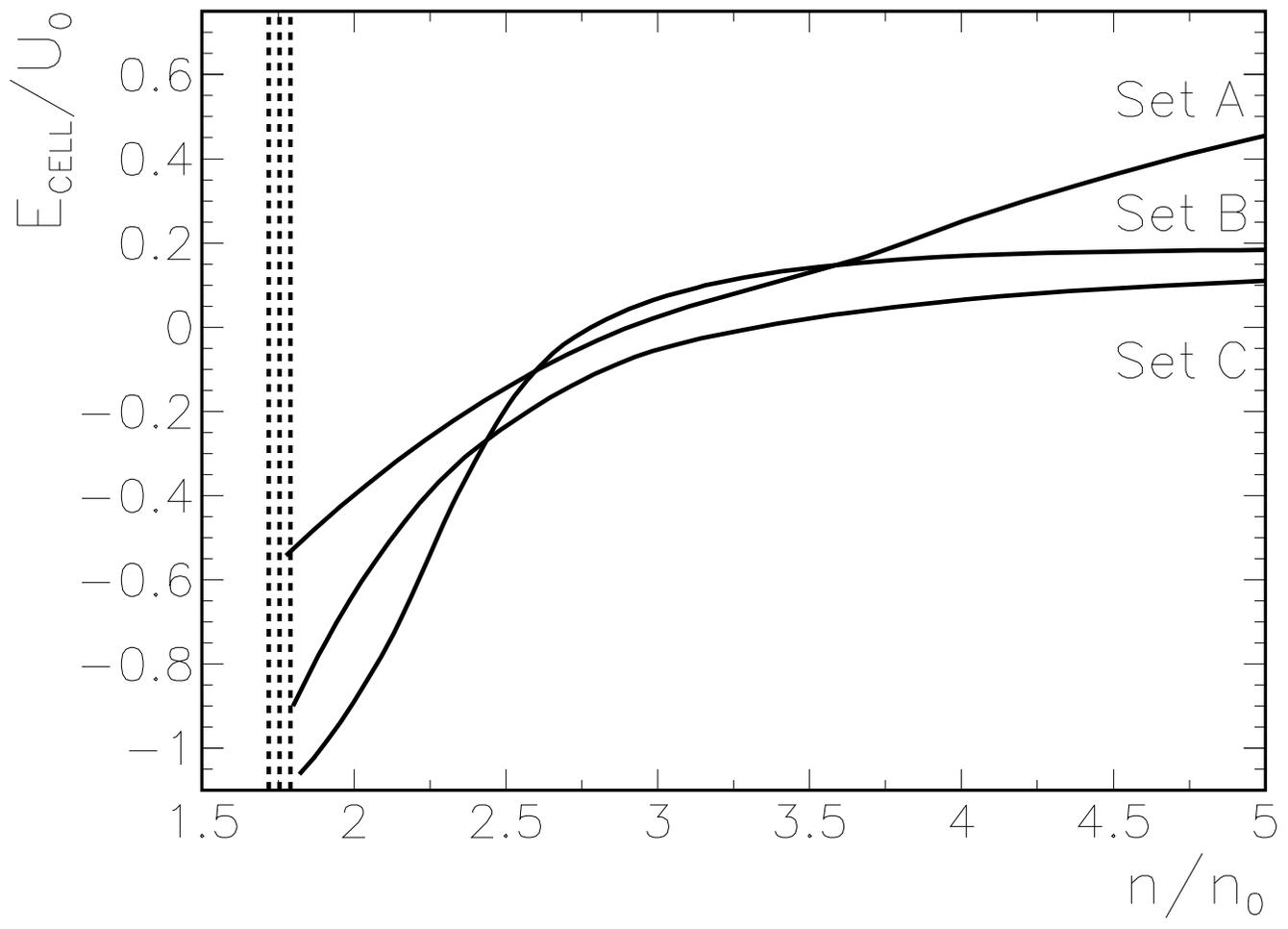
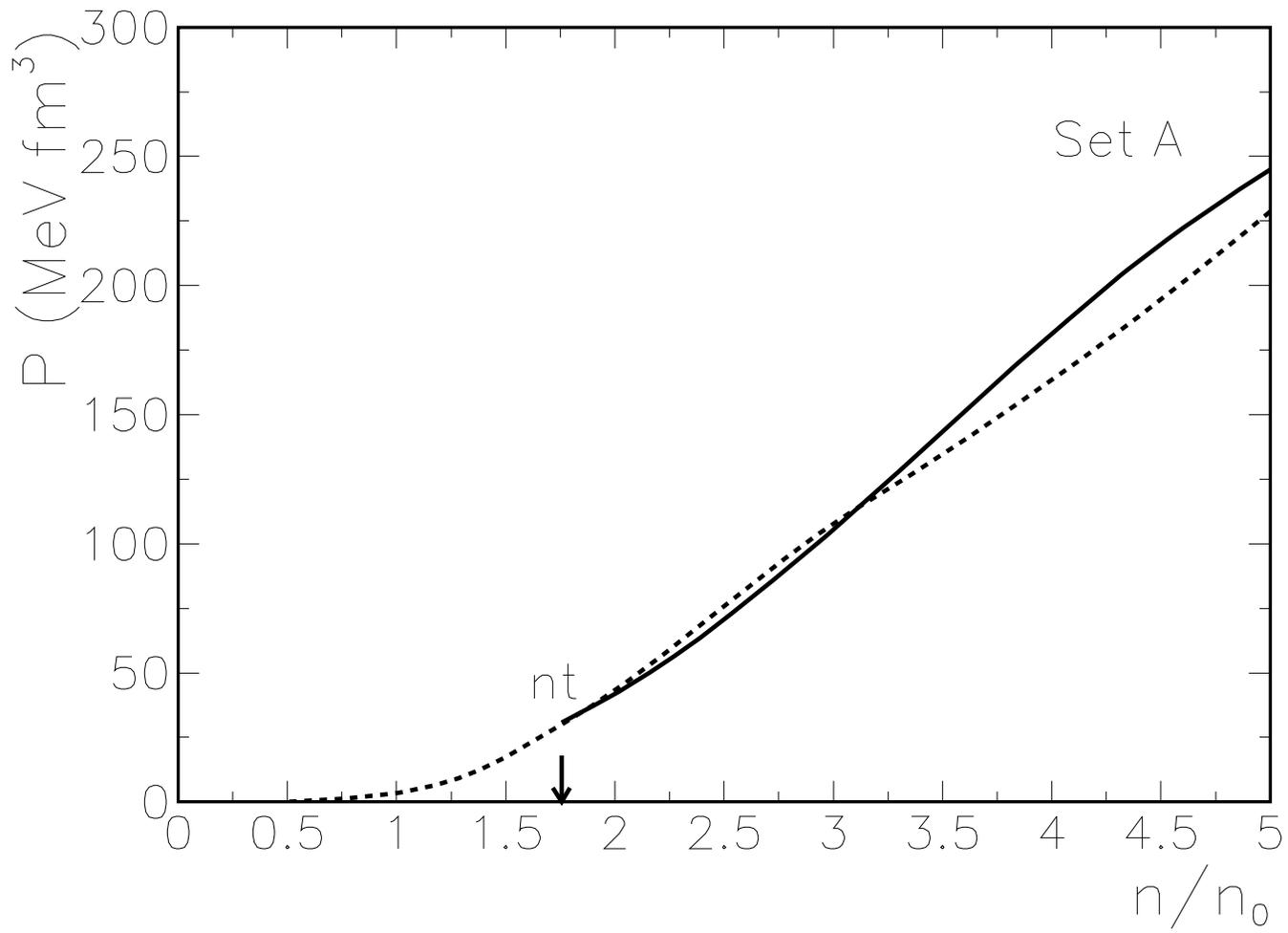
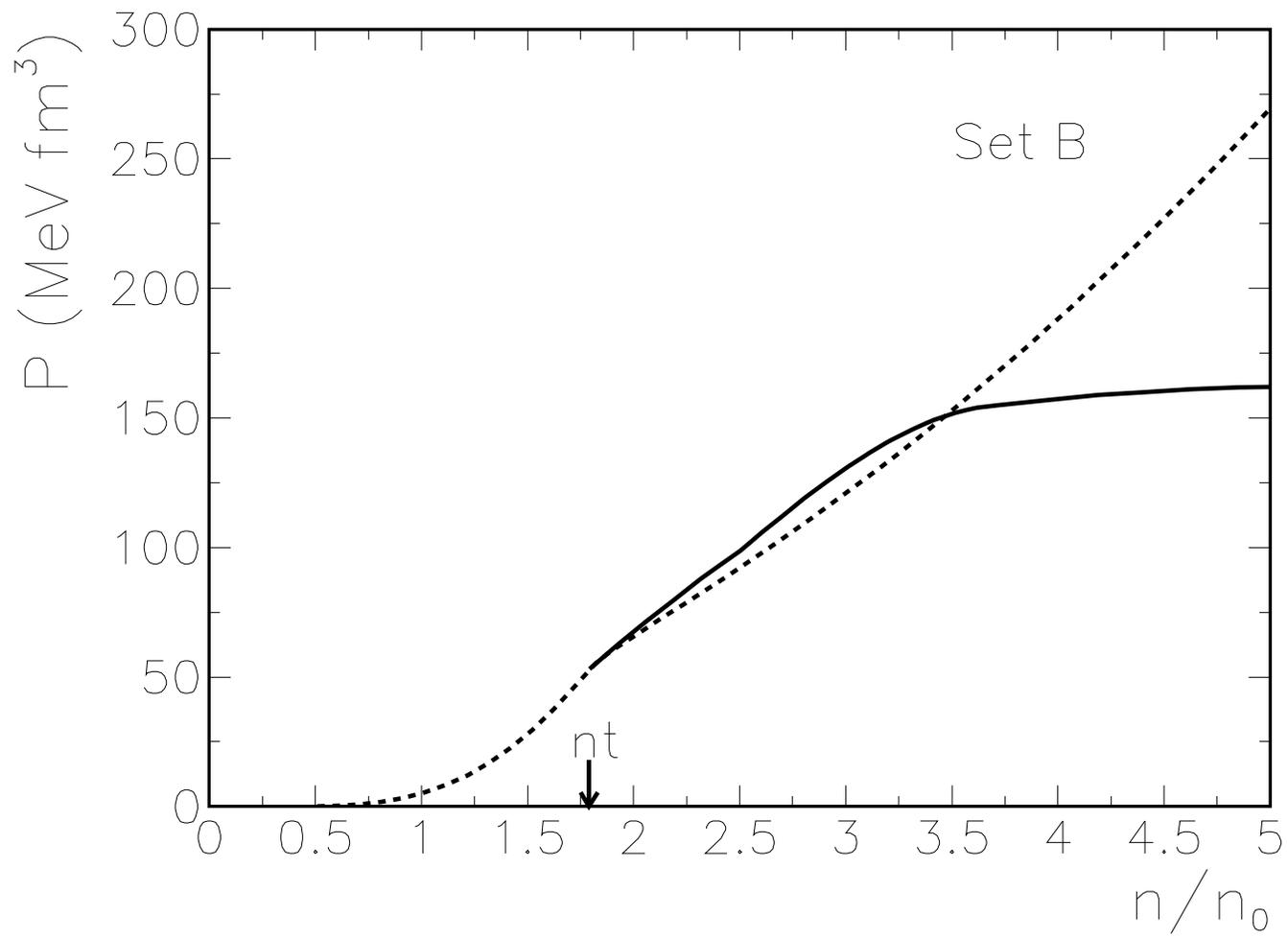
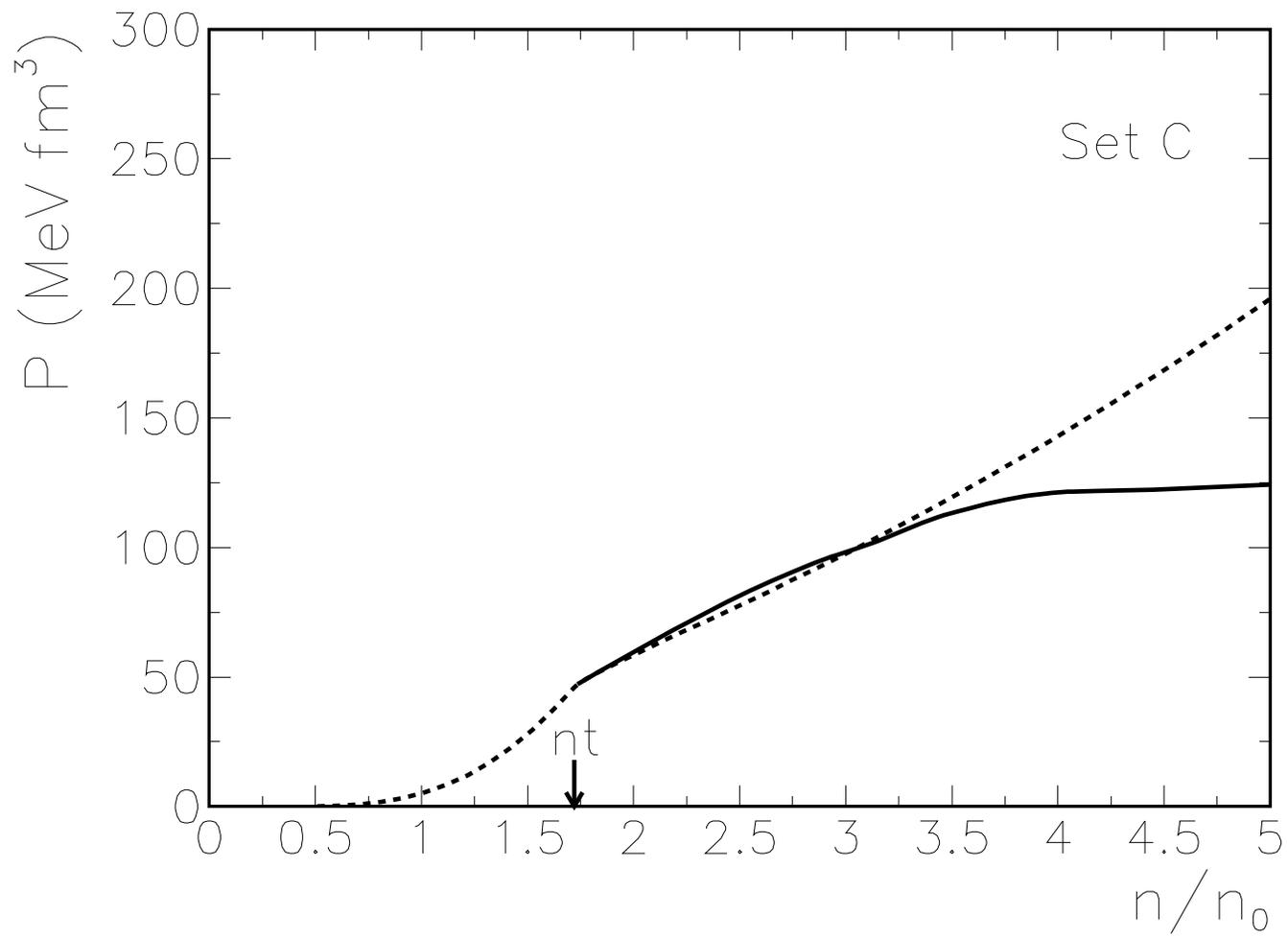
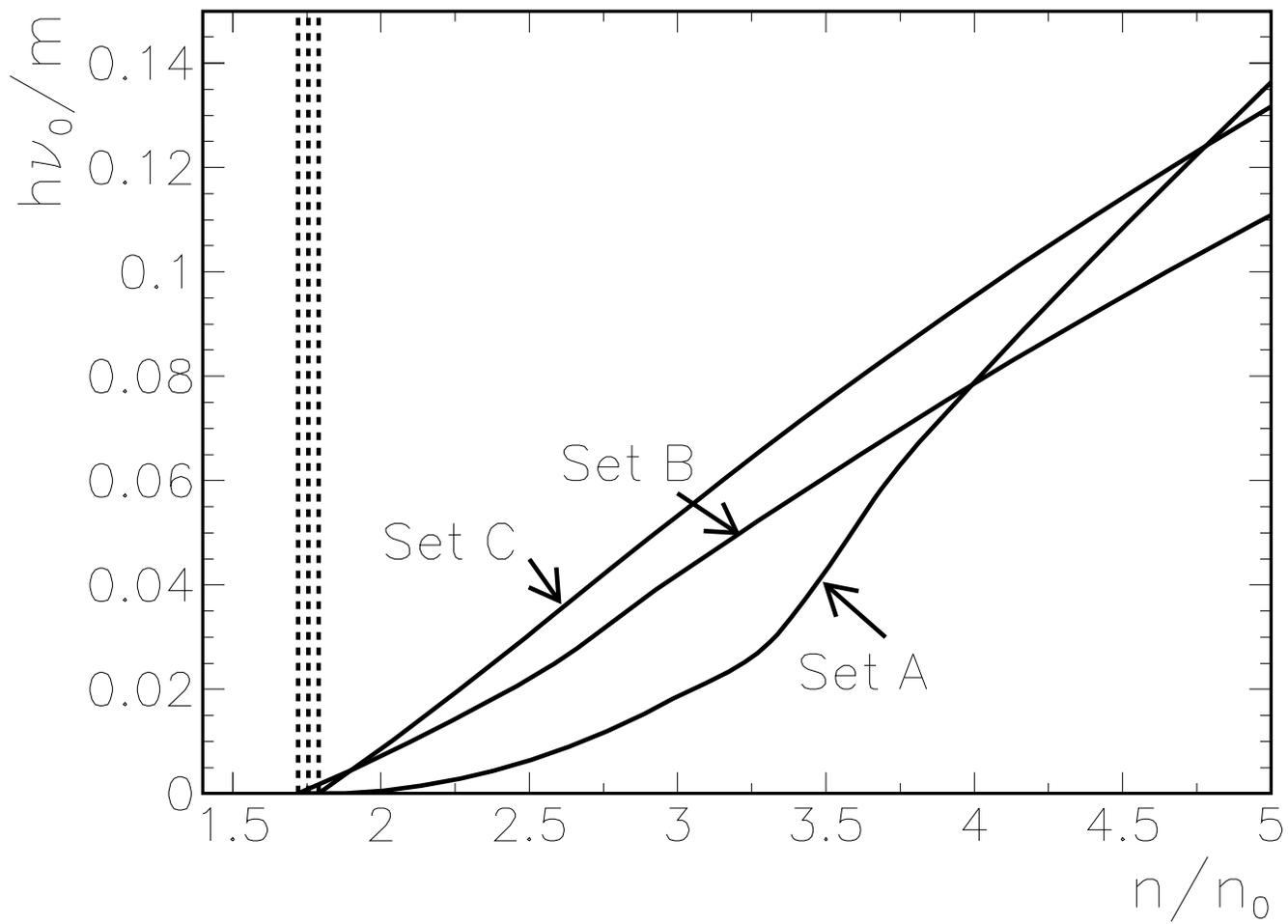
\end{document}